# Evolution of bilateral and multilateral collaboration of EU-14 countries across disciplines, 2010–2024

**Corresponding author:**
Myroslava Hladchenko
Centre for R&D Monitoring (ECOOM), University of Antwerp, Antwerp, Belgium
hladchenkom@gmail.com, myroslava.hladchenko@uantwerp.be

**Abstract**
This study explores the evolution of bilateral and multilateral research collaboration of nine EU-14 member states, both within Europe and globally, across six disciplines, between 2010 and 2024, using OpenAlex data. Results indicate that bilateral collaboration rates remained relatively stable and predominantly concentrated within EU-14 countries, followed by the USA, the UK, and China. Multilateral collaboration rates increased significantly across all disciplines, with the highest increase observed in medicine and the highest overall rates maintained in physics & astronomy. The same trend across disciplines was observed for the Relative Intensity of Collaboration (RIC). This reflects the growing importance of large-scale international research consortia in infrastructure-intensive fields that address global scientific challenges. RIC has increased for both bilateral and multilateral collaboration, with stronger growth in multilateral collaboration. Multilateral RIC fell below the expected level most frequently with South Korea, India, and China. Across both collaboration types, increases in collaboration rates were generally associated with increases in RIC.

No substantial changes in collaboration rates or RIC with the UK were observed following Brexit. A decline in multilateral collaboration with Russia in physics and astronomy coincided with its suspension from Horizon Europe in 2022, while the collaboration rate in medicine increased.

## 1.INTRODUCTION

The main idea behind the European Union's research and innovation strategy is that Europe's long-term technological and economic competitiveness relies on an integrated innovation system supported by robust collaborative networks and the widespread diffusion of knowledge across continents (European Commission, 2020; 2024). Examining collaboration within the EU-14 and between the EU-14 and global partners provides important insights into how the national research systems of EU-14 member states are embedded in European and global scientific dynamics (Melin, 2000; Beaver, 2001; Persson et al., 2004). Understanding the intensity and structure of these cross-border linkages is therefore crucial for evidence-based policymaking, research capacity building, and the further development of the European Research Area (Makkonen & Mitze, 2016).

Over the past decades, the evolution of scientific collaboration in Europe has been shaped by political integration and by both national and supranational research policies. Funding instruments at the EU and national levels have encouraged international collaboration. Successive EU enlargements, which granted new EU member states automatic paticipation in the EU Framework Programmes, together with the participation of non-EU countries through Horizon Association Agreements and low- and middle-income countries as partners, have contributed to closer international scientific integration (Glänzel & Schubert, 2001; Luukkonen et al., 1992).

Although bilateral collaboration still constitutes the dominant form of international co-authorship (Adams & Gurney, 2018) across disciplines (Glänzel & Schubert, 2004; Wagner & Leydesdorff, 2005; Gazni, Sugimoto & Didegah, 2012; Adams & Gurney, 2018; Aksnes &

Sivertsen, 2023) in part because it is easier to coordinate and is reinforced by geographical, cultural, linguistic and socioeconomic proximity (Hoekman et al., 2010; Scherngell & Barber, 2009; Chang & Huang, 2016), the rise of multilateral collaboration has become a defining feature of recent decades (Adams & Szomszor, 2024). Its expansion reflects the increasing requirement to address complex global challenges through large-scale projects (Adams & Szomszor, 2022; Hsiehchen et al., 2015; Lima-Toivanen et al., 2025; Burke, 2025) involving substantial investment, specialised infrastructure, and interdisciplinary teams (Adams, 2013). The EU Framework Programmes have been central to this shift, as their design prioritises transnational consortia and provides incentives for both intra-European and international participation (Veugelers, 2021).

However, disciplines differ significantly in their levels of international collaboration (Larivière et al., 2015; Cronin et al., 2003; Franceschet & Costantini, 2010; Guzmán-Valenzuela, 2023; Wang & Zha, 2025; Chang & Huang, 2016; Peng & Shi, 2024), and countries vary in their strategic disciplinary research priorities (Okamura, 2023; Gómez-Espés et al., 2024).

This study investigates the evolution of bilateral and multilateral research collaboration for nine EU-14 member states: Austria, Belgium, Finland, France, Germany, the Netherlands, Italy, Spain, and Sweden, within Europe and globally between 2010 and 2024, including potential effects of major geopolitical disruptions such as Brexit and the Russo-Ukrainian war. Collaboration partners include EU-14, EU-13, EU candidate countries, and major global research actors such as Australia, Brazil, Canada, Chile, China, Japan, India, Norway, South Africa, South Korea, Russia, Switzerland, the UK, and the USA. The analysis focuses on changes in the percentage of jointly co-authored articles and in the Relative Intensity of Collaboration (RIC) (Fuchs et al., 2021; Cao et al., 2025) across six disciplines: computer science, engineering, life & environmental sciences, medicine, physics and astronomy, and social sciences & humanities. The dataset is derived from OpenAlex (Gómez-Espés et al., 2024; Okamura, 2023) and consists of research articles published between 2010 and 2024 with at least one citation.

The novelty of this study lies in its integrated analysis of bilateral and multilateral scientific collaboration across disciplines, partner-country groups, and geopolitical contexts over the 2010–2024 period. Previous scientometric studies have typically examined either overall international collaboration or have focused on bilateral and multilateral collaboration in specific country contexts (Wagner & Leydesdorff, 2005; Adams, 2013; Adams & Gurney, 2018; Woolley et al., 2017). In contrast, this study systematically compares bilateral and multilateral collaboration simultaneously across multiple disciplines, partner regions, and European country groups over an extended period, while integrating collaboration rates and Relative Intensity of Collaboration (RIC) into a unified comparative framework.

## 2.LITERATURE REVIEW AND THEORETICAL BACKGROUND

### 2.1.Bilateral and multilateral collaboration

International scientific collaboration is shaped by a wide range of factors, including individual preferences of scholars as well as socioeconomic, political, geographical, cognitive, linguistic, and cultural proximities that foster cooperation (Hoekman et al., 2009; Glänzel & Schubert, 2004; Zitt et al., 2000; European Commission, 2025; Kwiek, 2021; Tu, 2024). Countries tend to collaborate most frequently with partners that are similar in terms of scientific capacity and economic development (Hoekman et al., 2009). As a result, core scientific countries tend to collaborate more intensively with one another than with countries outside the scientific core (Hoekman et al., 2009). On the other hand, R&D funding policies and national research capacity—including investment in infrastructure—strongly influence both the intensity and direction of collaboration (European Commission, 2025; Kwiek, 2021). These structural and policy factors shape the configuration of international collaboration, which can take either bilateral or multilateral forms.

Bilateral collaboration dominates international scientific publishing (Adams & Gurney, 2018) because coordinating research between two partners is organisationally simpler and remains the most common form of cross-border co-authorship (Glänzel & Schubert, 2004; Wagner & Leydesdorff, 2005; Gazni, Sugimoto & Didegah, 2012; Adams & Gurney, 2018). Bilateral links often persist even between non-neighbouring countries when partners share comparable socioeconomic conditions, as illustrated in the case of China and Canada (Wang & Zha, 2025). However, despite its quantitative dominance, bilateral collaboration increasingly coexists with rapidly expanding multilateral networks (Okamura, 2023; Adams & Gurney, 2018).

Multilateral collaboration has grown in response to global scientific challenges, enabling governments to overcome financial constraints through shared access to advanced research infrastructures (Lima-Toivanen et al., 2025; Burke, 2025) and facilitating the concentration of intellectual resources. Policy frameworks emphasise its strategic importance: the European Commission explicitly prioritises multilateral scientific collaboration (COM, 2021), and global collaboration is widely recognised as a driver of national research impact (Adams & Szomszor, 2024).

The distinction between bilateral and multilateral collaboration is theoretically important because the two forms reflect different organisational structures of knowledge production. Bilateral collaboration is often associated with stable, selective, and path-dependent partnerships shaped by geographic, linguistic, and historical proximity. In contrast, multilateral collaboration reflects increasingly networked forms of 'big science' characterised by distributed infrastructures, large consortia, interdisciplinary coordination, and broader international participation. Multilateral networks may facilitate wider diffusion of knowledge, integration of peripheral research systems, and access to shared scientific resources, although they may also reinforce inequalities by concentrating participation around highly connected scientific hubs.

### 2.2. Collaboration across disciplines

Disciplines differ substantially in their international collaboration patterns (Larivière et al., 2015; Cronin et al., 2003; Franceschet & Costantini, 2010; Guzmán-Valenzuela). According to Wagner (2005), collaboration often arises in response to the lack of equipment, materials, or specialised facilities. This observation is consistent with numerous disciplinary studies showing that collaboration is more prevalent where research is resource-intensive and requires shared infrastructure. As a result, collaboration is more common in the natural sciences than in the social sciences (Moody, 2004). Life sciences, medicine and physical sciences typically exhibit higher percentages of international co-authorship than social sciences (Ginza et al., 2012). International collaboration is particularly frequent in resource-based disciplines such as microbiology, immunology, molecular biology, and biochemistry, whereas theory-based disciplines such as economics, business, and the humanities show lower levels (Ginza et al., 2012). Mathematics tends to be less multi-authored overall, while chemistry and biomedical research display extensive multi-authorship (Glänzel, 2002).

Physics and medicine are not only the disciplines with the highest levels of international collaboration; they are also characterised by hyperauthorship (Cronin, 2001). Contemporary physics and biomedical research routinely involve large, multinational consortia. In experimental high-energy physics, the scale and cost of major research infrastructures necessitate the concentration of resources in a few laboratories (e.g., CERN, Fermilab, SLAC). This model of large-scale collaboration, however, raises concerns about authorship practices and accountability. Hyperauthorship can obscure individual contributions and challenge the credibility of the scientific communication system (Birnholtz, 2006). Rennie et al. (1997) suggested replacing lists of authors with lists of contributors, arguing that hyperauthorship masks intellectual input and complicates the attribution of credit and responsibility.

Hyperauthorship also has implications for research metrics. It can distort evaluations of research performance at national and organisational levels (Nogrady, 2023) and may bias funding allocation by inflating output indicators for institutions or researchers with limited actual contribution (Mryglod & Mryglod, 2020). These distortions are especially relevant for bibliometric indicators, which underlie national research assessment systems, and thus warrant careful consideration when analysing international collaboration trends.

### 2.3.EU Framework Programmes: EU integration and global collaboration

The scientific integration of European states was stimulated by the launch of the European Framework Programme in 1984 by the European Communities (EC). The subsequent EU enlargements of 2004 and 2007 further expanded the scope of scientific collaboration to include the joining Central and Eastern European countries (European Commission, 2015). Importantly, the Framework Programmes have strengthened collaboration not only within the EU but also globally.

EU member states participate automatically in Horizon Europe, whereas non-members must sign a Horizon Association Agreement to take part under equivalent conditions. Low- and middle-income countries may participate as partners and can receive EU funding, while high-income non-associated countries usually participate on a self-funded basis. Norway and Switzerland are associated with Horizon Europe. Despite Brexit, the UK remained associated with Horizon 2020 during the transition period (2016–2020) and was subsequently granted association status to Horizon Europe.

Several high-income countries—such as Australia, Brazil, Canada, Chile, China, Japan and the United States—are not associated with Horizon Europe. However, bilateral, jointly funded calls exist between Horizon Europe and countries such as Brazil, Japan, and China (Annette, 2025; European Commission, Directorate-General for Research & Innovation, n.d.).

South Korea became associated with Horizon Europe in 2025. India participates as an international partner and is eligible for EU funding. Although not fully associated, collaboration continues to deepen: in 2025, the EU and India launched joint calls worth approximately €60 million (Euraxess, 2024).

The political context also matters. In 2022, the European Commission suspended Russian public bodies from participation in Horizon 2020 and Horizon Europe, terminating their involvement in ongoing grants and halting further payments (EC, 2022). At the end of 2025, the European Commission announced the exclusion of Chinese universities, including the so-called “Seven Sons of National Defence,” from participating in roughly half of Horizon Europe calls (Matthews, 2025).

### 2.4.Dymanics of global scientific collaboration

From a global perspective, the USA occupies a central position in the worldwide scientific network (Olechnicka et al., 2019; Gui et al., 2019; Marini & Mouritzen, 2025), serving as the largest international collaborative partner for countries including China, the UK, Germany, France, Italy, Spain, the Netherlands (Kwiek, 2019) and South Korea (Gueye et al., 2022). Overall, the USA, U.K., Germany, France, Italy, and Canada are the core countries in the global collaboration network, collectively accounting for 82% of multinational publications. However, they are not the most collaborative countries. The most collaborative are Switzerland, Belgium, Denmark and Austria (Gazni et al., 2012).

Since the first decade of the 21st century, alongside the USA, China has gained a prominent position as the central hub of the global scientific collaboration network **(**Veugelers, 2017; Tang, 2024**).** Germany, the UK, and the Netherlands are among China’s top European collaborators (Gómez-Espés et al, 2024; Wang et al., 2017). Scientific collaboration between the European Union (EU) and Latin America has expanded steadily over the past two decades, driven by shared priorities in research and innovation (Belli & Nenoff, 2022) and the EU science diplomacy policy (Uribe-Mallarino, 2022). Spain and Portugal have historically been the strongest partners of Latin America due to linguistic ties and historical connections (Chinchilla-Rodríguez et al., 2016; Russell et al., 2020; Lemarchand, 2012; Uribe-Mallarino, 2022; Gueye et al., 2022).

## 3.METHODS

### 3.1. Relative Intensity of Collaboration

RIC is a widely used metric for quantifying the strength of research collaborations (Coccia & Bozeman, 2016; Fuchs et al., 2021; Luukkonen et al., 1992). This study uses RIC as the primary measure to assess the intensity of bilateral and multilateral international collaborations, allowing us to systematically compare EU state members collaborative activity across different partner countries and groups.

RIC (Fuchs et al., 2021) was used as a main metric in this study. The RIC indicator compares the share of country Y within the collaboration profile of country X to the share of Y within all collaborations in the whole network. RIC is calculated as the ratio of the observed share of co-authored publications between two countries to the expected share based on their total publication outputs (Fig 1). The RIC indicator is asymmetric and increases in value if collaboration between two countries increases. In this study, the publication-based interpretation of the RIC indicator developed by Fuchs et al., (2021) was used. RIC values of 1.0 indicates that the collaboration between two countries occurs as frequently as expected, based on their total publication outputs and the overall network structure, essentially reflecting random mixing. A value greater than 1.0 suggests preferential or stronger-than-expected collaboration, while a value below 1 indicates weaker-than-expected ties. This approach aligns with established practices in bibliometric studies of international co-authorship (Katz & Martin, 1997).

$$RIC(X,Y) := \left(\frac{C_{XY}}{C_X}\right) / \left(\frac{C_Y - C_{XY}}{T - C_X}\right) = \frac{C_{XY} \cdot (T - C_X)}{C_X \cdot (C_Y - C_{XY})}$$

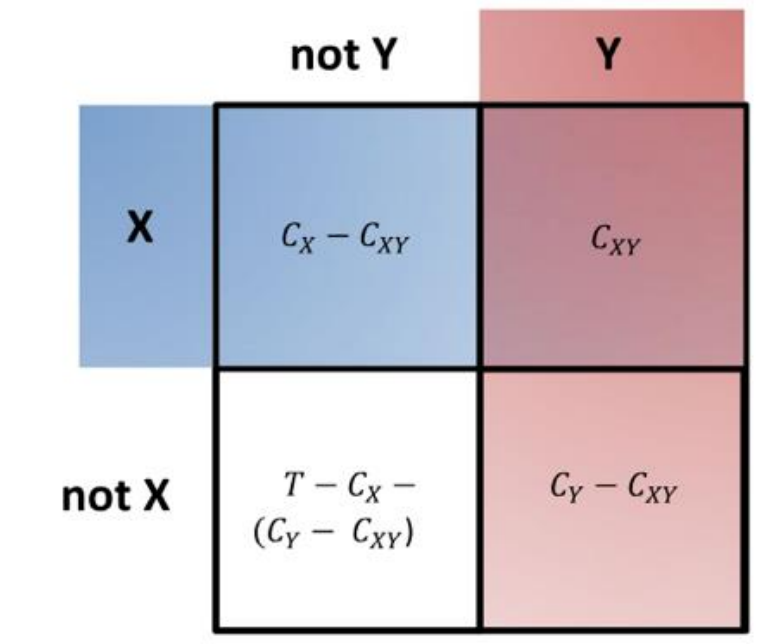

Fig. 1. Diagram and formular of RIC (Fuchs et a., 2021).

In the RIC formula:

- Cxy — the number of collaborative publications between entities X and Y;
- Cx — the total number of publications involving entity X;
- Cy — the total number of publications involving entity Y;
- T — the total number of publications in the analyzed network.

The collaboration rate was calculated as the proportion of publications involving international collaboration relative to the total publication output of the focal country (or institution/discipline) within a given disciplines and year. The denominator includes all publications associated with the focal entity, regardless of whether they were domestic or internationally co-authored. The numerator differs by collaboration type: for bilateral collaboration, it includes publications involving the focal entity and exactly one foreign country; for multilateral collaboration, it includes publications involving the focal entity and two or more foreign countries.

### 3.2. Data

The countries in question in this study are nine EU-14 member states: Austria, Belgium, Finland, France, Germany, Italy, Netherlands, Spain, Sweden. The collaborative partners include the two groups of countries. First, the EU groups: EU-14, EU-13 (Bulgaria, Croatia, Cyprus, Czech Republic, Estonia, Hungary, Latvia, Lithuania, Malta, Poland, Romania, Slovakia, Slovenia), and the EU candidates (Albania, Bosnia and Herzegovina, Georgia, Moldova, North Macedonia, Serbia, Turkey, Ukraine). Second, the global partners: Australia, Brazil, Canada, Chile, China, India, Japan, Norway, Russia, South Africa, South Korea, Switzerland, the UK, and the USA.

Data for this study were extracted from the OpenAlex dataset on BigQuery (Gómez-Espés et al., 2024; Okamura, 2023) stored by the InSySPo project, covering articles published between 2000 and 2025. For OpenAlex, only articles with at least one citation were included.

To focus on visible scientific publications and collaboration links, the analysis was restricted to articles that had received at least one citation at the time of data extraction. This approach also mitigates potential distortions associated with incomplete affiliation metadata in OpenAlex. Hladchenko (2026) revealed that around 30% of authors per year have incomplete affiliation metadata in OpenAlex, with this percentage rising substantially in recent years. The approach of including only cited articles from OpenAlex follows Engels et al. (2025). In addition, Hladchenko (2026a) showed that RIC values calculated from cited-only OpenAlex articles and from full-coverage Scopus articles capture highly similar temporal and structural trends in European collaboration networks. We nevertheless acknowledge that restricting the dataset to cited articles may affect coverage of the most recent publication years, as recently indexed articles are less likely to have accumulated citations and complete affiliation metadata at the time of data extraction. However, because the study examines collaboration structures rather than citation impact, citations were used only as a visibility and data-completeness threshold rather than as an evaluative metric.

This study explores collaboration across six disciplines: computer science, engineering, life & environmental sciences, medicine, physics & astronomy, and social sciences & humanities. The selection aimed to include fields with distinct epistemic cultures and, consequently, different collaboration patterns (Becher & Trowler, 2001; Cetina, 2009). The grouping is grounded in the formal OpenAlex classification framework of research subfields and existing bibliometric literature that recommends refinement of journal-level subject categories for more meaningful analysis (Boyack & Klavans, 2011; Glänzel & Schubert, 2003). This approach mirrors previous reclassification practices, such as Milojević (2020), who restructured subject categories into 14 fields in Web of Science. The aggregation of OpenAlex subfields into broader disciplinary categories is presented in Table 1 in the Appendix.

### 3.3. Statistical tests

Collaboration rate and RIC were calculated for each of nine explored EU-14 countries with each collaboration partner for bilateral and multilateral collaboration across six disciplines. Calculation was done in OpenAlex data on BigQuery stored by InSySPo project. Further analysis and visualisation was done in R.

To examine trends in scientific collaboration and ensure robustness, two linear mixed-effects models were fitted with country as a random intercept to account for country-specific variation. Model 1 was fitted to collaboration rates, while Model 2 was fitted to RIC. Both models included publication year (centered, reference: 2010), collaboration type (reference: bilateral), collaboration group (reference: EU-14), and discipline (reference: computer science), along with all possible interactions among these predictors. Predicted values for 2010 and 2024 were obtained using the emmeans package, allowing estimation of temporal changes in collaboration percentage (delta_per) and RIC (delta_RIC) for each combination of type, collaboration partner, and discipline. Following the modeling, a post-hoc analysis was conducted to explore the relationship between changes in collaboration magnitude and intensity over time. This included identifying cases where increases in collaboration percentage were accompanied by decreases in RIC, or vice versa, and visualising these patterns with scatter plots, linear trend lines, and faceting by collaboration type. This approach enables a comprehensive assessment of how collaboration patterns evolve across disciplines, collaboration types, and international partnerships, capturing both the magnitude and intensity of scientific collaboration.

## 4.RESULTS

### 4.1.Total number of bilateral and multilateral articles across disciplines

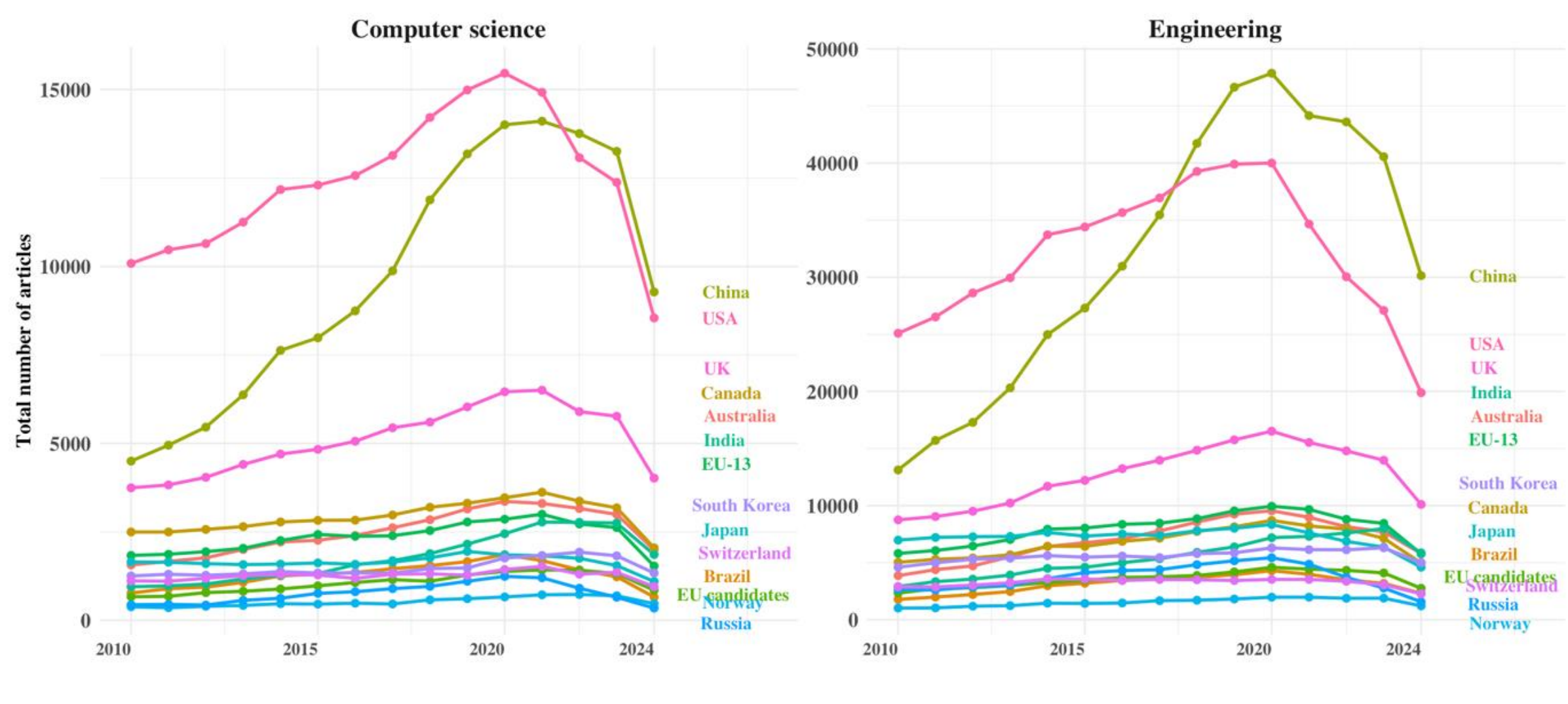

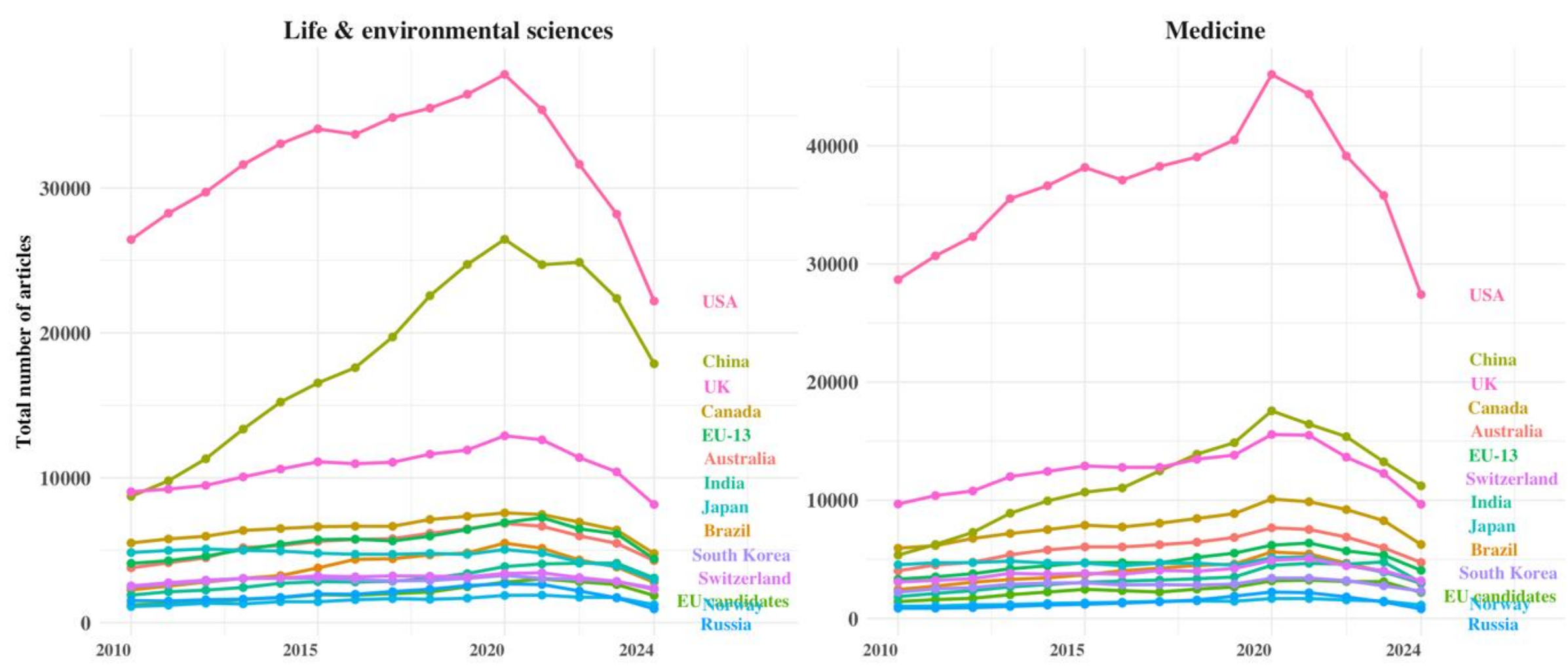

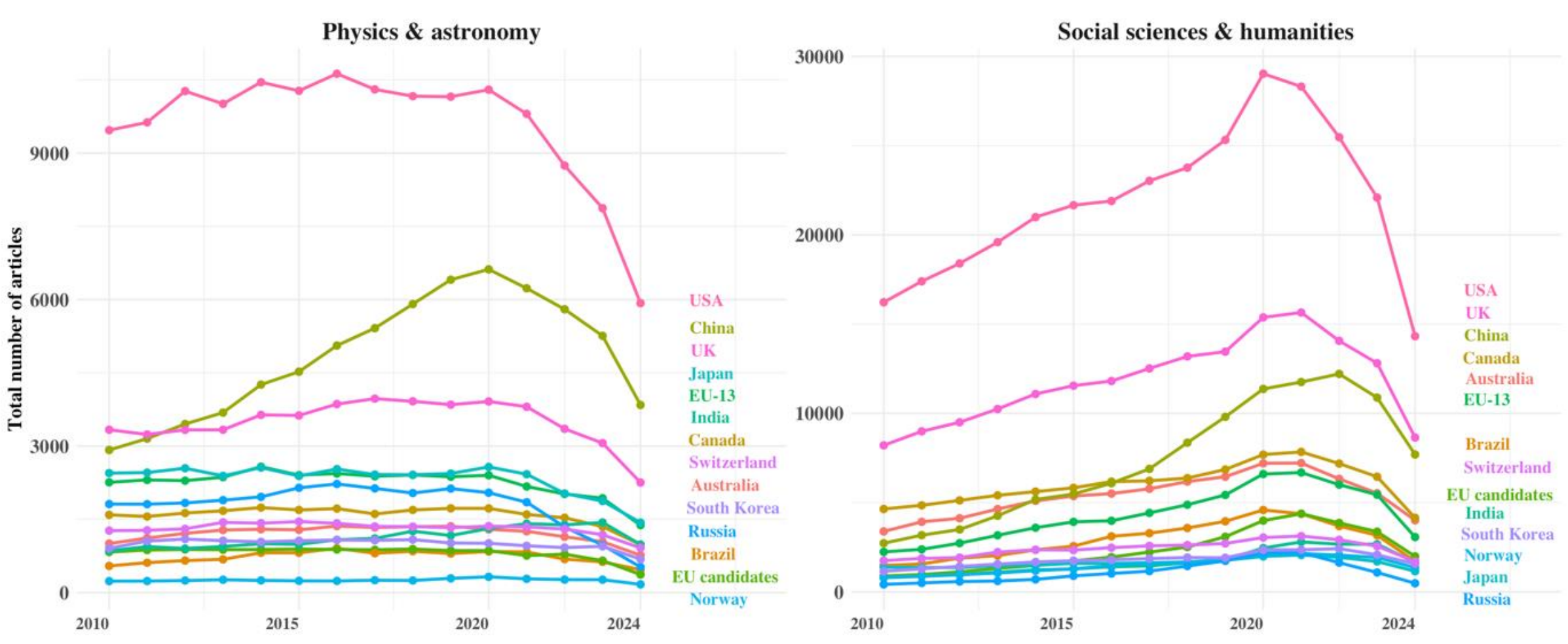

**Fig. 2.** Total number of internationally co-authored articles (bilateral collaboration).

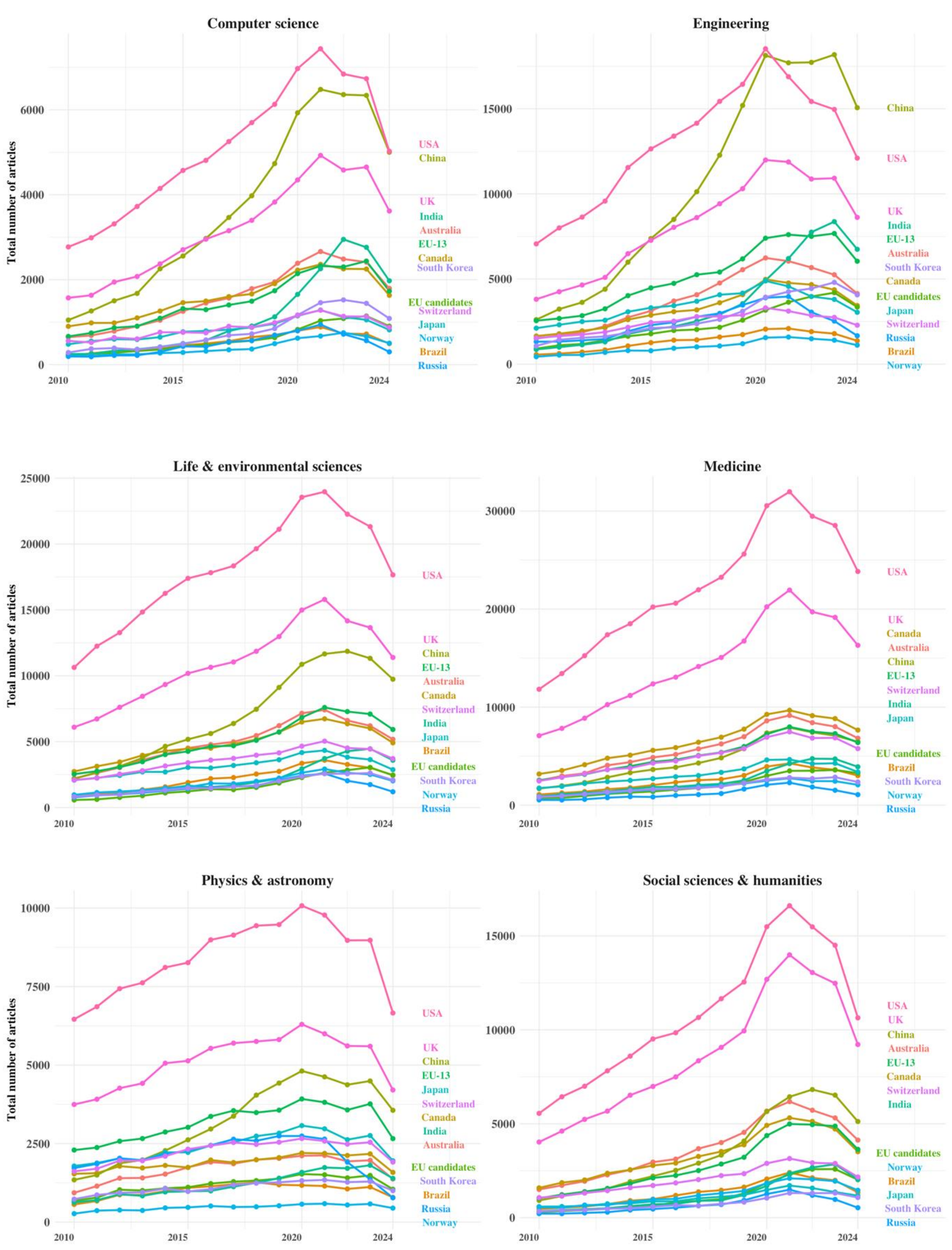

**Fig. 3.** Total number of internationally co-authored articles (multilateral collaboration).

Figures 2 and 3 illustrate the total number of cited internationally co-authored articles in bilateral and multilateral collaborations across partner countries. In bilateral collaboration, the USA dominated across most disciplines. The main exceptions were computer science and engineering, where China dominated in recent years. However, the USA continued to substantially outperform China in medicine and social sciences & humanities. For multilateral collaboration, China surpassed the USA in engineering and largely caught up in computer science. In the remaining disciplines, however, China continued to lag behind both the USA and the UK, with the largest differences observed in medicine and social sciences & humanities. The decline observed in recent years is likely attributable to the inclusion of only cited articles in the dataset.

### 4.2.Evolution of collaboration rate from 2010 to 2024

A Model 1, linear mixed-effects model with country as a random intercept was fitted to collaboration percentages by collaboration type, discipline, and collaboration partner, controlling for discipline (reference: computer science), collaboration partner (reference: EU-14), collaboration type (reference: bilateral), and publication year (reference: 2010) (Fig. 4). The model explained nearly all variation in collaboration percentages ($R^2_m = 0.972$; $R^2_c = 0.973$), indicating that temporal, disciplinary, and partner-specific factors accounted for substantially more variation than unobserved country-level heterogeneity.

The results revealed diverging trajectories for bilateral and multilateral collaboration. Whereas bilateral collaboration remained relatively stable between 2010 and 2024 (5.98% → 5.78%), multilateral collaboration increased significantly (15.6% → 17.9%). This increase was observed across all disciplines, although its magnitude varied considerably. The strongest growth occurred in medicine (+4.3%) and life & environmental sciences (+2.9%), whereas engineering and computer science exhibited comparatively weaker increases. Physics & astronomy consistently displayed the highest levels of multilateral collaboration, reflecting reliance on large-scale international infrastructures and consortium-based research. By contrast, computer science and engineering exhibited lower multilateral collaboration rate.

A major structural shift concerned the increasing collaboration rate with China. Across both collaboration types, China emerged as the fastest-growing partner, particularly in engineering, computer science, physics & astronomy, and life & environmental sciences. In bilateral collaboration, the most pronounced increase occurred in engineering, where collaboration with China rose from 3.4% to 14.8%. In physics & astronomy, the multilateral collaboration rate with China nearly doubled (13.1% → 25.9%).

At the same time, both bilateral and multilateral collaboration with the USA declined across most disciplines. The sharpest bilateral decreases were observed in social sciences & humanities (20.7% → 11.5%), medicine (24.8% → 16.1%), and life & environmental sciences (19.9% → 11.6%). Nevertheless, the USA remained among the most ctitical collaboration partners across disciplines, particularly in computer science and physics & astronomy.

The analysis also demonstrates differentiated patterns of European scientific integration. EU-14 countries remained the dominant collaboration partners across all disciplines, accounting for approximately 35–37% of bilateral collaboration and ~60-70% of multilateral collaboration by 2024. Bilateral collaboration with EU-13 (4-5% across disciplines) and EU candidate countries (~1.1-1.4%) showed little change over time, whereas multilateral collaboration increased slightly across most disciplines. This pattern suggests that scientific integration of EU-13 and EU candidate countries occurred primarily through multilateral research networks rather than through bilateral partnerships.

Collaboration with Russia exhibited a substantial decline, although the magnitude of the decline varied across disciplines and collaboration types.The strongest reductions occurred in physics & astronomy, where bilateral collaboration declined from 5.9% to 3.5% and multilateral collaboration from 24.3% to 14.8%. Engineering also showed marked decreases, particularly in multilateral collaboration. In contrast, medicine and life & environmental sciences displayed comparatively limited changes.

Among other countries, a small rise was observed in multilateral collaboration across most disciplines. With Japan, the highest growth was observed in multilateral collaboration in medicine and with India in computer science.

The top 10 increases and decreases in collaboration rates across collaboration partners and disciplines are reported in the Appendix (Tables 2 and 3).

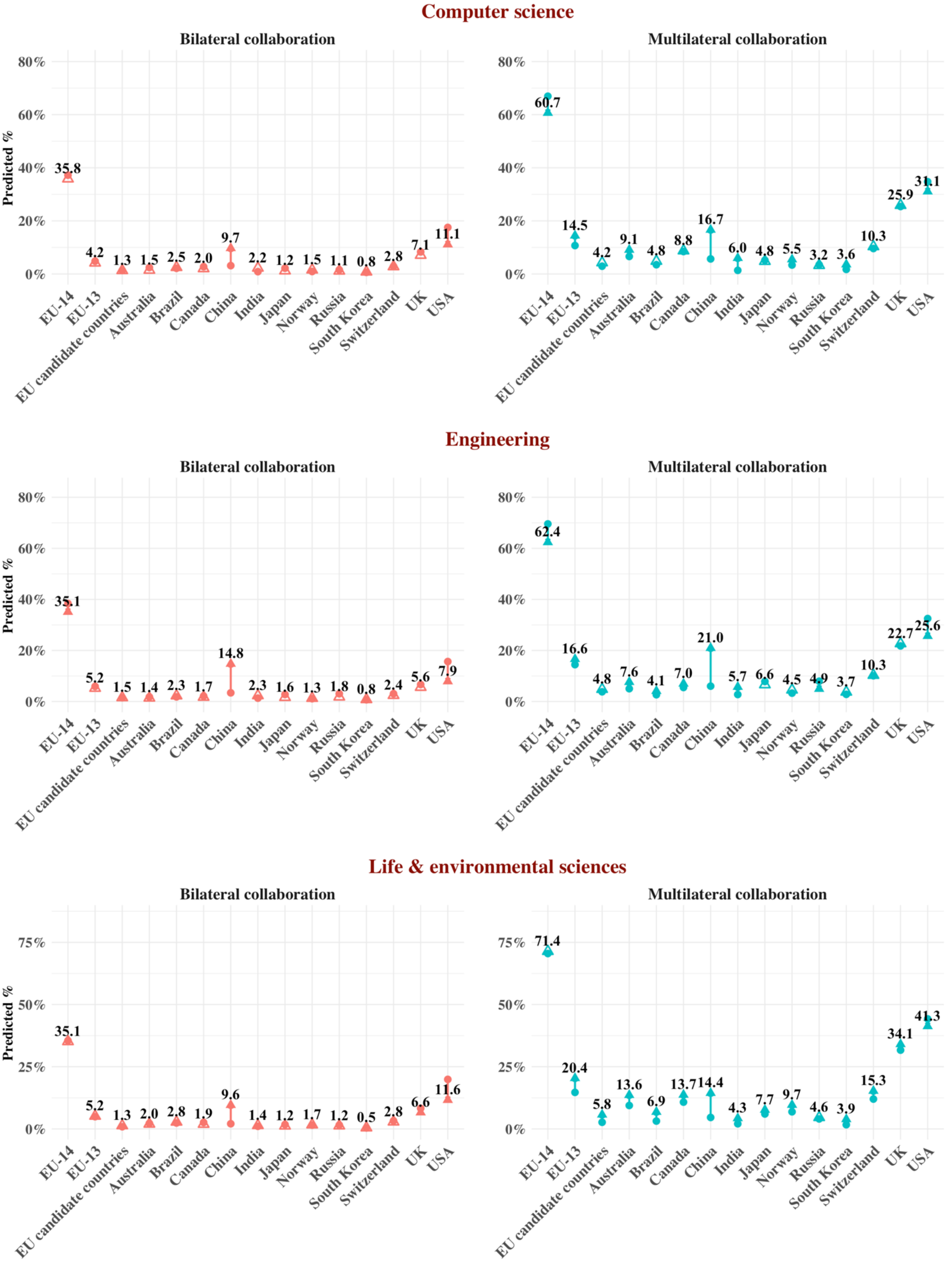

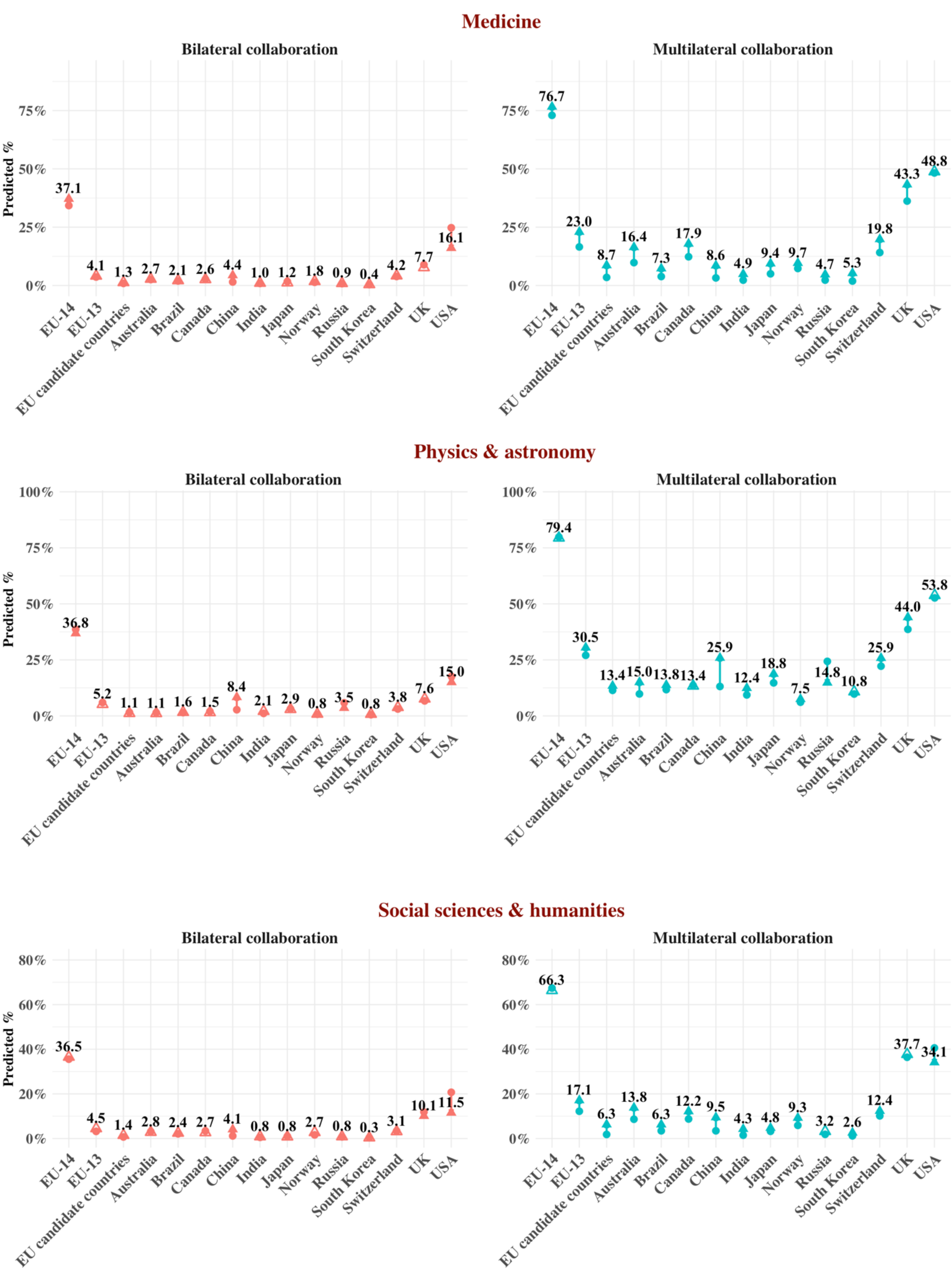

**Fig. 4.** Predicted percentage of jointly co-authored articles across disciplines, 2010-2024.

## 4.3. Collaboration rate in 2024

Figure 5 shows that in 2024, the bilateral collaboration rate was below 5% with most partners and across disciplines. Higher bilateral collaboration rates were primarily observed within the EU-14 (≈ 26–50%) and, to a lesser extent, with the USA, the UK, and China. However, substantial disciplinary variation existed. China occupied a particularly prominent position in computer science, engineering, and physics & astronomy, especially with Finland (16%; 22.9%; 13.8%) and Sweden (16.1%; 24.6%; 10.2%), whereas collaboration with China was substantially lower in medicine and social sciences & humanities. These patterns indicate that collaboration with China was strongly concentrated in STEM disciplines. The highest bilateral collaboration rates with the USA were observed for Germany, the Netherlands, and Italy.

Across all collaboration pairs and disciplines, multilateral collaboration rates exceeded bilateral ones, confirming the growing importance of broader international research networks. The highest levels of multilateral collaboration were observed in physics & astronomy and medicine, consistent with the disciplinary patterns identified above. Multilateral collaboration was particularly concentrated within the EU-14 (approximately 55–86%), reflecting the continued centrality of intra-European scientific integration. Collaboration with EU-13 and EU candidate countries displayed clear disciplinary differentiation, with the highest rates observed in medicine and physics & astronomy.

High levels of multilateral collaboration were also observed with the UK, the USA, and China, frequently exceeding 10%. Collaboration with Australia and Canada exceeded 10% in several disciplines, particularly life & environmental sciences, medicine, social sciences & humanities, and physics & astronomy. By contrast, collaboration with Brazil, India, Japan, and South Korea was generally lower, exceeding 10% only in physics & astronomy and, for India, also in medicine.

Several country pairs exhibited comparatively high collaboration rates across disciplines. In particular, multilateral collaboration with China was particularly high for Finland and Sweden, while Austria, Germany, and Italy showed relatively high bilateral collaboration with Switzerland. Strong collaboration was also observed among the Nordic countries, especially between Norway, Finland, and Sweden. Spain exhibited comparatively high collaboration with Brazil relative to other EU-14 countries, reflecting linguistic, historical, and regional ties that continued to shape collaboration structures.

**Computer science**

| | Bilateral collaboration | | | | | | | | | Multilateral collaboration | | | | | | | | |
|---|---|---|---|---|---|---|---|---|---|---|---|---|---|---|---|---|---|---|
| | Austria | Belgium | Finland | France | Germany | Italy | Netherlands | Spain | Sweden | Austria | Belgium | Finland | France | Germany | Italy | Netherlands | Spain | Sweden |
| EU-14 | 52.1 | 46.7 | 33.8 | 26.8 | 31.1 | 38.3 | 40.1 | 33.1 | 34.4 | 76 | 72.6 | 53.6 | 56.9 | 59.5 | 62.3 | 71.5 | 58.4 | 61.1 |
| EU-13 | 6.1 | 2.7 | 4.2 | 4 | 3.9 | 3.9 | 2.9 | 4 | 2.5 | 22.6 | 17.4 | 14.9 | 13.3 | 13.1 | 15.7 | 12.9 | 17.4 | 11.2 |
| EU candidate countries | 1.8 | 0.8 | 2.5 | 1.1 | 1.3 | 1.5 | 1.5 | 2.2 | 1.8 | 6.4 | 6.1 | 6.3 | 4.2 | 4 | 5.5 | 3.4 | 6.3 | 5.5 |
| Australia | 1.5 | 0.8 | 1.6 | 1.1 | 1.5 | 1.3 | 2.6 | 1.3 | 1.6 | 7.6 | 9.3 | 9.8 | 7.4 | 9.1 | 8.1 | 11.4 | 8.9 | 12.1 |
| Brazil | 0.7 | 2.5 | 0.7 | 2.5 | 1 | 1.6 | 1.5 | 2.3 | 1.7 | 6.1 | 4.6 | 4.4 | 5.6 | 4.4 | 5.2 | 5.1 | 6.9 | 3.9 |
| Canada | 1.5 | 1.7 | 1.2 | 3.8 | 2.7 | 1.6 | 2.2 | 1.2 | 3.1 | 8 | 10.9 | 9.6 | 9.8 | 10.2 | 8.7 | 10.6 | 7.9 | 8 |
| China | 4.7 | 9.8 | 16 | 10.3 | 9.8 | 9.7 | 11.9 | 7 | 16.1 | 11.1 | 13.3 | 20.7 | 16.9 | 17.1 | 14.7 | 13 | 16 | 17.9 |
| India | 1.7 | 1.9 | 4.2 | 2.6 | 1.9 | 3 | 1.3 | 2 | 3.4 | 6.9 | 5.7 | 8.4 | 6.9 | 6 | 9.1 | 5.1 | 7.4 | 9.2 |
| Japan | 1.8 | 0.8 | 0.9 | 2.1 | 1.7 | 0.9 | 1 | 0.6 | 0.9 | 5.2 | 5.4 | 4.2 | 5.6 | 5.5 | 5.4 | 5.8 | 5.4 | 4.4 |
| Norway | 1 | 1 | 3.2 | 1.2 | 1.2 | 1 | 1.2 | 1.2 | 2.6 | 6.4 | 6.3 | 8.8 | 3.9 | 4.8 | 5.1 | 7.9 | 4.5 | 10.1 |
| Russia | 1 | 0.2 | 0.2 | 0.6 | 0.5 | 0.3 | 0.3 | 0.5 | 0.4 | 1.6 | 2.8 | 1.7 | 2.1 | 1.8 | 2.1 | 1 | 1.5 | 1.7 |
| South Korea | 0.3 | 0.6 | 1.2 | 0.7 | 1 | 0.8 | 0.4 | 0.8 | 1.5 | 4.5 | 3.7 | 6.3 | 3.5 | 3.2 | 3.5 | 2.3 | 6.5 | 4.1 |
| Switzerland | 3.2 | 2.5 | 1.1 | 3.6 | 5.5 | 4.2 | 3 | 1.7 | 1.1 | 15.1 | 15.2 | 7.7 | 11.2 | 14.6 | 12.1 | 13 | 9.9 | 9.8 |
| UK | 3.5 | 8 | 5.8 | 4.9 | 8.8 | 9.1 | 8.3 | 5.7 | 6.7 | 25.2 | 28.7 | 27.2 | 24.9 | 27.7 | 26.4 | 32.7 | 23.8 | 28.9 |
| USA | 11.8 | 9.4 | 10.4 | 12.1 | 18.2 | 12.9 | 13.6 | 9.3 | 8.6 | 34.2 | 34.6 | 33.1 | 30.9 | 36.9 | 31.9 | 37.1 | 28.8 | 32.7 |

**Engineering**

| | Bilateral collaboration | | | | | | | | | Multilateral collaboration | | | | | | | | |
|---|---|---|---|---|---|---|---|---|---|---|---|---|---|---|---|---|---|---|
| | Austria | Belgium | Finland | France | Germany | Italy | Netherlands | Spain | Sweden | Austria | Belgium | Finland | France | Germany | Italy | Netherlands | Spain | Sweden |
| EU-14 | 48.2 | 42.5 | 30.6 | 27.2 | 27.9 | 37.6 | 39.7 | 37.8 | 32.1 | 74.8 | 71.1 | 57.7 | 56.8 | 55.3 | 63.6 | 71.7 | 63.3 | 63.9 |
| EU-13 | 10 | 4.4 | 6.2 | 4 | 5.1 | 6.5 | 2.6 | 4 | 4 | 22.6 | 14.9 | 19.7 | 15.5 | 16.9 | 17.8 | 13.4 | 16.4 | 16.5 |
| EU candidate countries | 2.9 | 0.6 | 3.2 | 1 | 1.9 | 2.2 | 1.2 | 1.1 | 1.2 | 7.5 | 4.5 | 6.2 | 4.5 | 5.6 | 5.8 | 4.3 | 6.2 | 5.6 |
| Australia | 1 | 0.9 | 1.5 | 1.1 | 1.7 | 1.3 | 1.4 | 1.2 | 1 | 7.8 | 7.9 | 7.5 | 6.8 | 7.6 | 6.1 | 9 | 7 | 8.8 |
| Brazil | 1 | 1.5 | 1.5 | 2.8 | 1.9 | 1.7 | 1.6 | 4.2 | 2.2 | 3.9 | 4.2 | 4.8 | 3.5 | 3.2 | 4.3 | 3.7 | 5.6 | 3.8 |
| Canada | 0.6 | 1.7 | 1.9 | 3.3 | 2 | 1.6 | 1.1 | 1 | 1.4 | 7.6 | 6.6 | 9 | 7.3 | 6.9 | 7.1 | 8.3 | 5.6 | 7 |
| China | 8.6 | 18.1 | 22.9 | 14.9 | 17.5 | 10 | 17.2 | 8.1 | 24.6 | 14.9 | 21.7 | 23.2 | 20.4 | 24.9 | 16.7 | 21.3 | 16.6 | 24 |
| India | 1.4 | 1.8 | 3.8 | 2.1 | 3.6 | 2.8 | 2.3 | 2.2 | 4 | 5.9 | 3.8 | 6.4 | 5.4 | 6.4 | 7.8 | 4.3 | 6.3 | 8 |
| Japan | 2.1 | 1.2 | 2.6 | 3.1 | 2.4 | 1.2 | 1.3 | 1.1 | 1.2 | 7 | 5.5 | 8.3 | 7.4 | 8.3 | 6.1 | 6.5 | 6.3 | 6.9 |
| Norway | 1.4 | 0.5 | 1.4 | 0.6 | 1.2 | 1 | 1.4 | 0.8 | 2.3 | 6 | 3 | 7.5 | 2.8 | 3.4 | 3.4 | 4.9 | 3.2 | 6.8 |
| Russia | 1.2 | 0.6 | 0.3 | 1.4 | 1.3 | 0.8 | 0.5 | 0.5 | 0.4 | 3.5 | 3.4 | 2.5 | 2.9 | 2.8 | 2.7 | 1.7 | 3.4 | 2.8 |
| South Korea | 0.6 | 1.1 | 1.2 | 1 | 1.4 | 0.4 | 0.4 | 0.5 | 1.1 | 3.4 | 4.7 | 5.1 | 3.7 | 4.2 | 4.3 | 3.5 | 4.1 | 4.7 |
| Switzerland | 2.7 | 2.2 | 0.8 | 2.6 | 4.1 | 3.7 | 2.7 | 1.8 | 1.5 | 14.6 | 11.2 | 10.2 | 10.6 | 12.8 | 10.9 | 12.4 | 9 | 10.2 |
| UK | 4.4 | 4.6 | 4.9 | 4 | 5.5 | 7.1 | 7.1 | 5.6 | 6 | 20.4 | 24.6 | 23.4 | 22.2 | 23.3 | 23 | 27.4 | 22.5 | 21.1 |
| USA | 7 | 7.3 | 5.1 | 9 | 12.6 | 10.8 | 11.3 | 7.3 | 7.2 | 26.9 | 24.8 | 24.2 | 27.1 | 30.5 | 27.6 | 31.2 | 24.5 | 25.8 |

## Life & Environmental Sciences

### Bilateral collaboration

| | Austria | Belgium | Finland | France | Germany | Italy | Netherlands | Spain | Sweden |
|---|---|---|---|---|---|---|---|---|---|
| EU-14 | 46.1 | 42.6 | 32.6 | 29.4 | 29 | 37 | 37.4 | 38.2 | 34.5 |
| EU-13 | 9.4 | 4.2 | 7.3 | 3.2 | 4.9 | 7 | 2.4 | 4 | 4 |
| EU candidate countries | 2.1 | 0.5 | 2.4 | 0.6 | 1.8 | 2.3 | 0.9 | 1.2 | 1.4 |
| Australia | 1.7 | 1.4 | 1.7 | 1.8 | 2.1 | 1.4 | 2.2 | 1.4 | 1.8 |
| Brazil | 0.7 | 1.5 | 1.2 | 2.9 | 2.1 | 2.1 | 1.6 | 4.6 | 1.8 |
| Canada | 1.1 | 1.8 | 1.7 | 4.7 | 1.9 | 1.6 | 2 | 1.2 | 1.8 |
| China | 5.1 | 13.3 | 16.7 | 7.7 | 11.6 | 4.8 | 13.4 | 4.6 | 14 |
| India | 0.9 | 0.9 | 2 | 1.6 | 2.2 | 1.5 | 1.1 | 0.7 | 2 |
| Japan | 1.2 | 0.9 | 1.3 | 2.1 | 1.3 | 1 | 0.9 | 0.9 | 1.2 |
| Norway | 0.8 | 0.7 | 2.5 | 1.1 | 1.4 | 0.8 | 1.6 | 1 | 4.6 |
| Russia | 0.8 | 0.9 | 1 | 1 | 0.7 | 0.6 | 0.4 | 0.2 | 0.7 |
| South Korea | 0.9 | 0.8 | 0.3 | 0.7 | 0.7 | 0.3 | 0.3 | 0.4 | 0.4 |
| Switzerland | 3.9 | 2.3 | 2.1 | 3.7 | 5.3 | 3.7 | 2.7 | 1.9 | 1.3 |
| UK | 5 | 5 | 5.9 | 6.1 | 6.6 | 8.2 | 7.5 | 7.2 | 7.6 |
| USA | 11.7 | 10.4 | 11.5 | 12.7 | 16.3 | 15.2 | 16 | 11.5 | 11.9 |

### Multilateral collaboration

| | Austria | Belgium | Finland | France | Germany | Italy | Netherlands | Spain | Sweden |
|---|---|---|---|---|---|---|---|---|---|
| EU-14 | 78.6 | 77.5 | 68.1 | 69.6 | 64.2 | 73 | 75.3 | 73.6 | 73 |
| EU-13 | 28.8 | 21.8 | 27.5 | 18 | 18.9 | 22.3 | 17.3 | 20.5 | 19.1 |
| EU candidate countries | 7.6 | 6.1 | 5.3 | 5.1 | 5.6 | 8.3 | 4.3 | 6.7 | 5 |
| Australia | 11.9 | 12.9 | 13.7 | 13 | 11.7 | 11.6 | 14.1 | 13.2 | 14 |
| Brazil | 5.9 | 6.6 | 6 | 7.6 | 6.2 | 6.8 | 6.8 | 9.7 | 6.5 |
| Canada | 13.4 | 14 | 14 | 15.6 | 12.8 | 13 | 14.5 | 13.3 | 13.8 |
| China | 11.5 | 13.7 | 15.9 | 14.4 | 16.2 | 12.3 | 14.5 | 12.8 | 16.3 |
| India | 4.7 | 4.4 | 5 | 4.4 | 4.7 | 5.3 | 3.8 | 4.2 | 5.5 |
| Japan | 8.6 | 8.2 | 8.1 | 8.7 | 7.6 | 7.5 | 6.9 | 7.1 | 6.4 |
| Norway | 9.3 | 7.8 | 18.1 | 7.5 | 7.5 | 6.7 | 8.9 | 7.6 | 15 |
| Russia | 3.6 | 3.1 | 4.7 | 2.9 | 3.4 | 3.1 | 2.3 | 2.8 | 3.4 |
| South Korea | 3.9 | 4.1 | 3.6 | 3.9 | 3.8 | 4.2 | 2.7 | 4.3 | 2.9 |
| Switzerland | 20.7 | 17.4 | 16 | 15.6 | 16 | 14.9 | 15.4 | 13.3 | 15.2 |
| UK | 31.7 | 34.7 | 34.1 | 33.3 | 31.7 | 33.4 | 40.4 | 33.7 | 36 |
| USA | 40.9 | 43.7 | 40 | 43.9 | 45 | 42 | 45.7 | 42.8 | 42.8 |

## Medicine

### Bilateral collaboration

| | Austria | Belgium | Finland | France | Germany | Italy | Netherlands | Spain | Sweden |
|---|---|---|---|---|---|---|---|---|---|
| EU-14 | 53.1 | 48.6 | 35.8 | 30.8 | 30.1 | 32.3 | 38.9 | 35.5 | 35.6 |
| EU-13 | 7.2 | 3.2 | 5.9 | 2.7 | 4.7 | 5.4 | 1.8 | 3.7 | 3.1 |
| EU candidate countries | 2.2 | 0.9 | 1.7 | 0.9 | 2 | 2.1 | 0.5 | 0.8 | 1.1 |
| Australia | 1.2 | 2.4 | 6.8 | 1.7 | 2.2 | 1.4 | 3.4 | 1.5 | 4.1 |
| Brazil | 0.9 | 1.6 | 0.8 | 2.2 | 1.4 | 1.9 | 1.8 | 3.9 | 1.4 |
| Canada | 1.6 | 2.4 | 1.6 | 6.2 | 2.1 | 2.3 | 2.1 | 1.8 | 2.6 |
| China | 2 | 5.2 | 7.8 | 4.2 | 6.4 | 2.3 | 4.5 | 1.8 | 8.1 |
| India | 0.2 | 0.7 | 1.2 | 1.1 | 1 | 1.5 | 0.7 | 1 | 1.4 |
| Japan | 0.8 | 1.5 | 2.1 | 1.6 | 1.3 | 0.9 | 0.8 | 0.8 | 1.1 |
| Norway | 0.6 | 0.2 | 3.8 | 0.4 | 0.8 | 0.4 | 1.5 | 1.1 | 6.9 |
| Russia | 0.9 | 0.6 | 0.8 | 0.5 | 0.6 | 0.7 | 0.1 | 0.3 | 0.6 |
| South Korea | 0.3 | 0.3 | 0.4 | 0.5 | 0.8 | 0.3 | 0.4 | 0.2 | 0.1 |
| Switzerland | 6.5 | 2.8 | 2.4 | 5.4 | 9.3 | 6 | 2.6 | 1.6 | 1.8 |
| UK | 4.1 | 6.1 | 5.8 | 5.9 | 6.8 | 10.5 | 9.2 | 7.7 | 7.4 |
| USA | 13.5 | 12.8 | 13.5 | 17.5 | 21.7 | 22.2 | 22.6 | 18.2 | 15.3 |

### Multilateral collaboration

| | Austria | Belgium | Finland | France | Germany | Italy | Netherlands | Spain | Sweden |
|---|---|---|---|---|---|---|---|---|---|
| EU-14 | 84.7 | 83 | 73.4 | 77.3 | 72.8 | 77.4 | 78.5 | 78.6 | 75.5 |
| EU-13 | 30 | 24.7 | 27.3 | 20.5 | 21.2 | 24.6 | 20.9 | 23 | 19.5 |
| EU candidate countries | 11.8 | 9.3 | 9.3 | 8.3 | 8 | 11.7 | 6.9 | 10.8 | 6.9 |
| Australia | 15.7 | 16.9 | 17.2 | 15.5 | 13.9 | 14.1 | 17.2 | 15.3 | 17 |
| Brazil | 6.9 | 7.1 | 5.8 | 7.4 | 6.2 | 7.8 | 6.5 | 10.1 | 5.9 |
| Canada | 19.4 | 17.9 | 15.1 | 20.8 | 18.1 | 17.9 | 19.2 | 18.3 | 15.3 |
| China | 6.4 | 8.1 | 8.6 | 10.2 | 9.9 | 9.1 | 7.7 | 9.1 | 9.9 |
| India | 5.8 | 4.8 | 5.6 | 5.1 | 4.9 | 6.7 | 4.4 | 5.4 | 5.1 |
| Japan | 9.8 | 10.1 | 7 | 11.6 | 9.4 | 10.1 | 7.8 | 10.7 | 6.8 |
| Norway | 8.5 | 7.5 | 19.6 | 6.6 | 6 | 6.4 | 8.4 | 6.7 | 16.3 |
| Russia | 4.2 | 3 | 3.9 | 3.5 | 3.1 | 4 | 2.2 | 3.6 | 2.7 |
| South Korea | 4.8 | 4.9 | 3.3 | 6.4 | 5.1 | 6.1 | 3.9 | 7.9 | 3.2 |
| Switzerland | 28.3 | 20.8 | 16.4 | 20.7 | 22.6 | 19.3 | 18.5 | 17.4 | 15.3 |
| UK | 39.2 | 44.1 | 41.2 | 42.6 | 39 | 41.7 | 47 | 43.6 | 45.7 |
| USA | 46.9 | 49.2 | 40.7 | 54.2 | 54 | 51 | 51.6 | 51.7 | 48.5 |

## Social sciences & humanities

### Bilateral collaboration

| | Austria | Belgium | Finland | France | Germany | Italy | Netherlands | Spain | Sweden |
|---|---|---|---|---|---|---|---|---|---|
| EU-14 | 48.1 | 49.5 | 35.7 | 29.8 | 34.4 | 38.2 | 39.5 | 31.3 | 34.2 |
| EU-13 | 6.7 | 3.4 | 6.5 | 2.3 | 4.1 | 5.8 | 2.6 | 3.3 | 2.8 |
| EU candidate countries | 1.4 | 0.9 | 1.6 | 1 | 1.8 | 2.3 | 1.1 | 1.1 | 1.5 |
| Australia | 2.5 | 3.1 | 3.5 | 1.8 | 3 | 1.6 | 4 | 1.6 | 4.1 |
| Brazil | 0.6 | 2.2 | 0.7 | 2.9 | 1.6 | 2.1 | 1.3 | 4.4 | 1 |
| Canada | 2.1 | 2.2 | 1.8 | 6.6 | 1.9 | 2.5 | 2 | 1.2 | 2.9 |
| China | 2.3 | 4.9 | 5.9 | 4.9 | 4.4 | 3.6 | 5.3 | 3.2 | 5.1 |
| India | 0.5 | 0.2 | 0.5 | 1.7 | 0.9 | 1.2 | 0.8 | 0.5 | 0.9 |
| Japan | 1 | 0.5 | 1.1 | 1.4 | 1 | 0.6 | 0.5 | 0.5 | 0.9 |
| Norway | 1.4 | 1.4 | 5.7 | 1.3 | 2.1 | 1.1 | 2 | 1.1 | 10.4 |
| Russia | 0.4 | | 0.4 | 0.4 | 0.3 | 0.5 | 0.1 | 0.3 | 0.1 |
| South Korea | 0.4 | 0.2 | 0.5 | 0.5 | 0.3 | 0.3 | 0.2 | 0.2 | 0.2 |
| Switzerland | 4.9 | 2.6 | 1.2 | 3.8 | 7.8 | 3 | 2.6 | 1.3 | 1.8 |
| UK | 8.3 | 8 | 10.9 | 8.8 | 9.6 | 13.6 | 11.3 | 8.9 | 10.5 |
| USA | 10.1 | 9.4 | 10.7 | 13.5 | 16.5 | 13.9 | 15 | 10.9 | 12.4 |

### Multilateral collaboration

| | Austria | Belgium | Finland | France | Germany | Italy | Netherlands | Spain | Sweden |
|---|---|---|---|---|---|---|---|---|---|
| EU-14 | 79.5 | 75.8 | 61 | 62.7 | 65.3 | 68.4 | 67.9 | 64.8 | 68.7 |
| EU-13 | 24.9 | 18.1 | 20.5 | 15.2 | 16.7 | 19.9 | 15 | 20.3 | 15.4 |
| EU candidate countries | 9.5 | 7.1 | 7.4 | 6 | 6 | 9.5 | 5 | 8.7 | 5.3 |
| Australia | 16.6 | 14.5 | 17.4 | 12.8 | 14.9 | 11.8 | 16.4 | 12.6 | 17.8 |
| Brazil | 6.7 | 5.9 | 5 | 5.8 | 5.1 | 6.7 | 4.9 | 11.4 | 4.3 |
| Canada | 13.3 | 12.3 | 9.9 | 14 | 13.2 | 13.2 | 13.5 | 11.6 | 13.4 |
| China | 7.5 | 9.6 | 10.5 | 11.3 | 9.8 | 9.1 | 9.3 | 9.6 | 9.8 |
| India | 4.6 | 5 | 5.2 | 7 | 4.3 | 6 | 4.1 | 4.5 | 5.3 |
| Japan | 6.4 | 5 | 5.7 | 5.5 | 5.9 | 5.8 | 5.3 | 5.2 | 5.2 |
| Norway | 9.6 | 9.6 | 17.3 | 5.8 | 7.6 | 7 | 9.4 | 6.7 | 18.9 |
| Russia | 2.9 | 1.8 | 2.5 | 2.2 | 1.8 | 2.8 | 1.2 | 2.6 | 1.6 |
| South Korea | 4 | 2.2 | 4.2 | 3.1 | 2.8 | 3.1 | 2.3 | 3.8 | 2.9 |
| Switzerland | 17.4 | 15.5 | 9.7 | 12.6 | 16.2 | 13.1 | 13.9 | 10.3 | 12.2 |
| UK | 37.8 | 40.4 | 37.4 | 37.3 | 37.7 | 39.2 | 41.4 | 34.6 | 39.2 |
| USA | 35.6 | 35.2 | 29.8 | 37.5 | 41.6 | 34.9 | 40.2 | 33.9 | 38.3 |

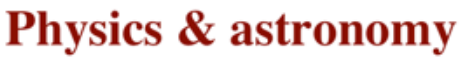

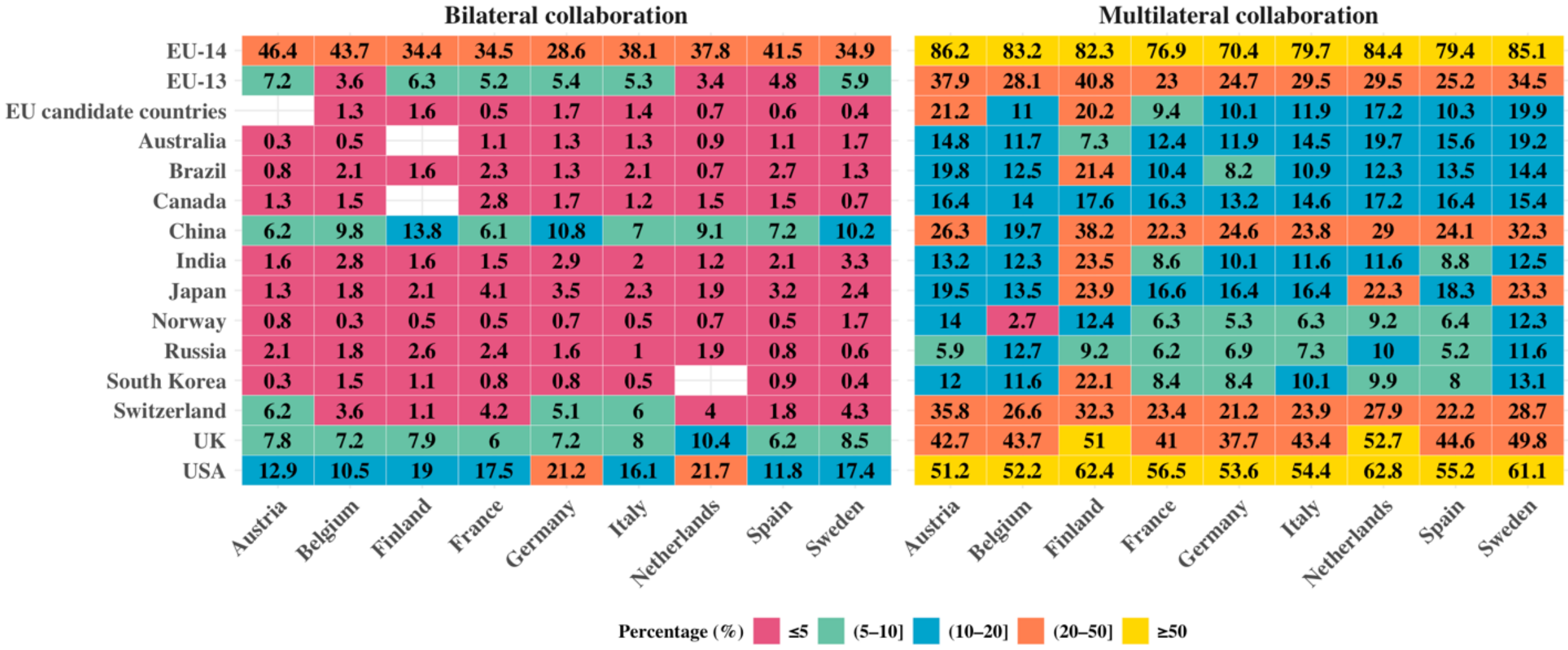

**Fig. 5.** Percentage of jointly co-authored articles in 2024.

## 4.4.Evolution of RIC from 2010 to 2024

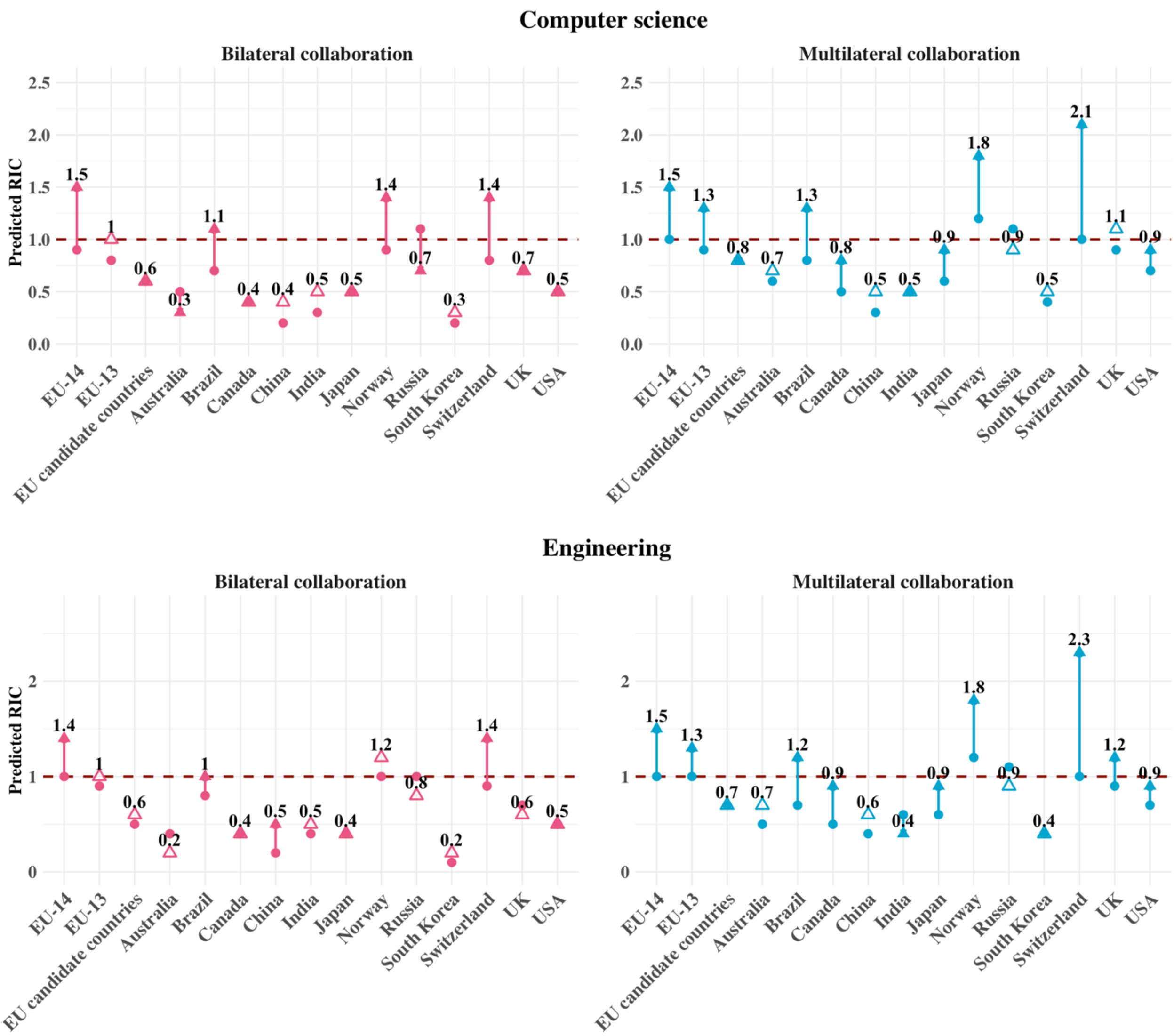

Computer science

Bilateral collaboration | Multilateral collaboration

Predicted RIC

EU-14, EU-13, EU candidate countries, Australia, Brazil, Canada, China, India, Japan, Norway, Russia, South Korea, Switzerland, UK, USA

Engineering

Bilateral collaboration | Multilateral collaboration

Predicted RIC

EU-14, EU-13, EU candidate countries, Australia, Brazil, Canada, China, India, Japan, Norway, Russia, South Korea, Switzerland, UK, USA

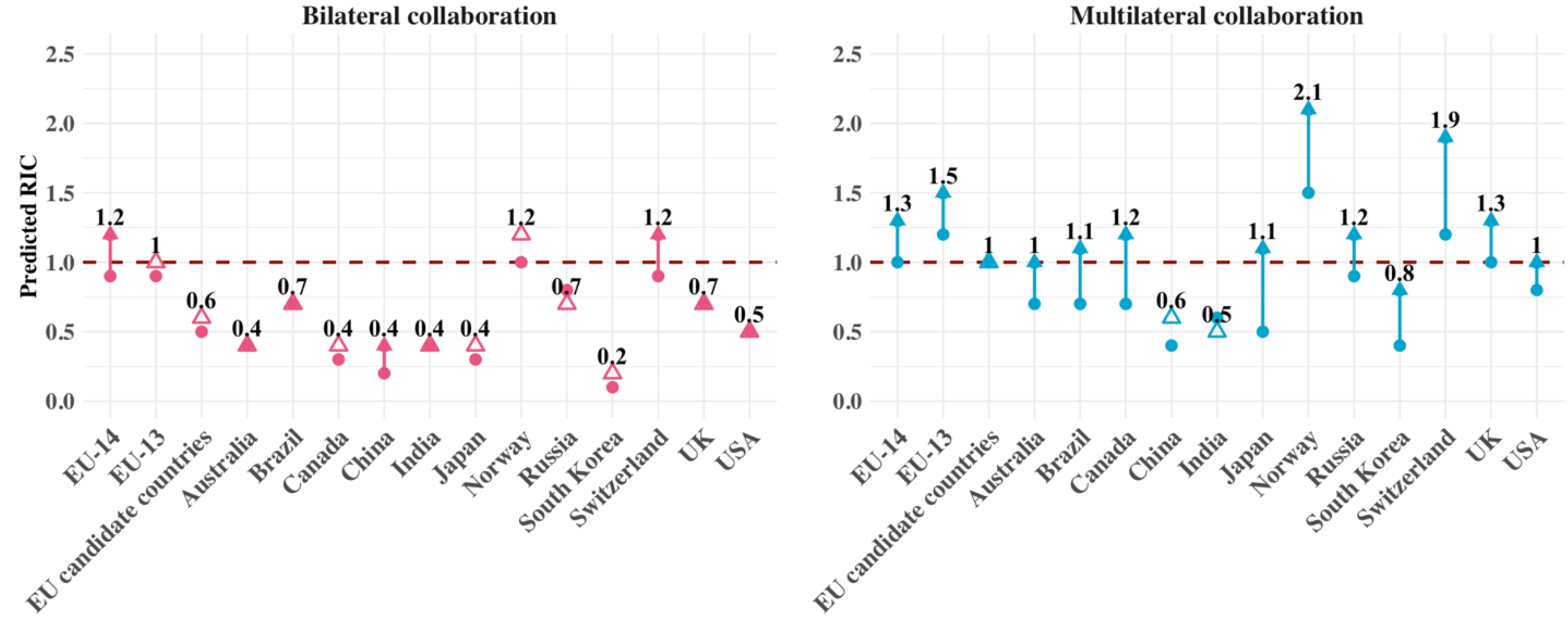

Life & environmental sciences
Bilateral collaboration
Multilateral collaboration
Predicted RIC
0.0
0.5
1.0
1.5
2.0
2.5
1.2
1
0.6
0.4
0.7
0.4
0.4
0.4
0.4
1.2
0.7
0.2
1.2
0.7
0.5
1.3
1.5
1
1
1.1
1.2
0.6
0.5
1.1
2.1
1.2
0.8
1.9
1.3
1
EU-14
EU-13
EU candidate countries
Australia
Brazil
Canada
China
India
Japan
Norway
Russia
South Korea
Switzerland
UK
USA

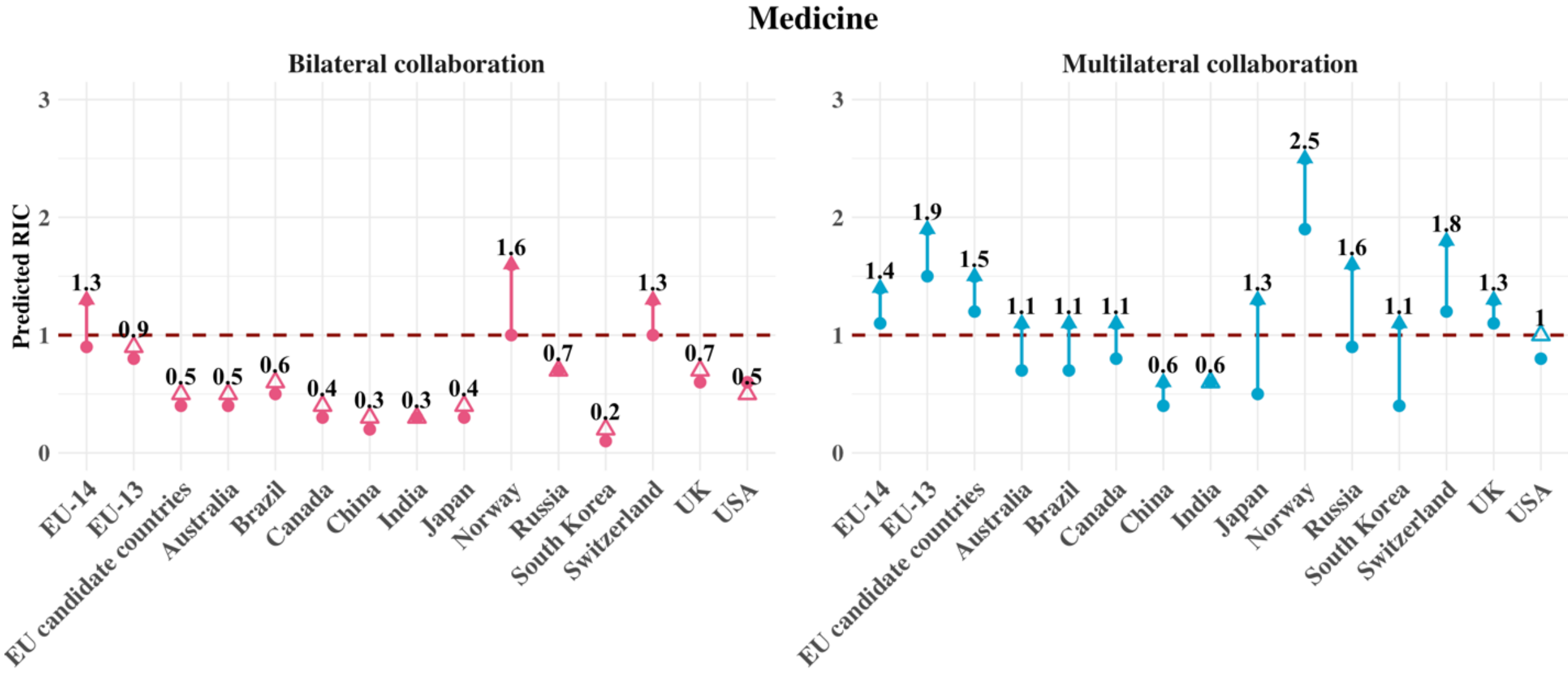

Medicine
Bilateral collaboration
Multilateral collaboration
Predicted RIC
0
1
2
3
1.3
0.9
0.5
0.5
0.6
0.4
0.3
0.3
0.4
1.6
0.7
0.2
1.3
0.7
0.5
1.4
1.9
1.5
1.1
1.1
1.1
0.6
0.6
1.3
2.5
1.6
1.1
1.8
1.3
1
EU-14
EU-13
EU candidate countries
Australia
Brazil
Canada
China
India
Japan
Norway
Russia
South Korea
Switzerland
UK
USA

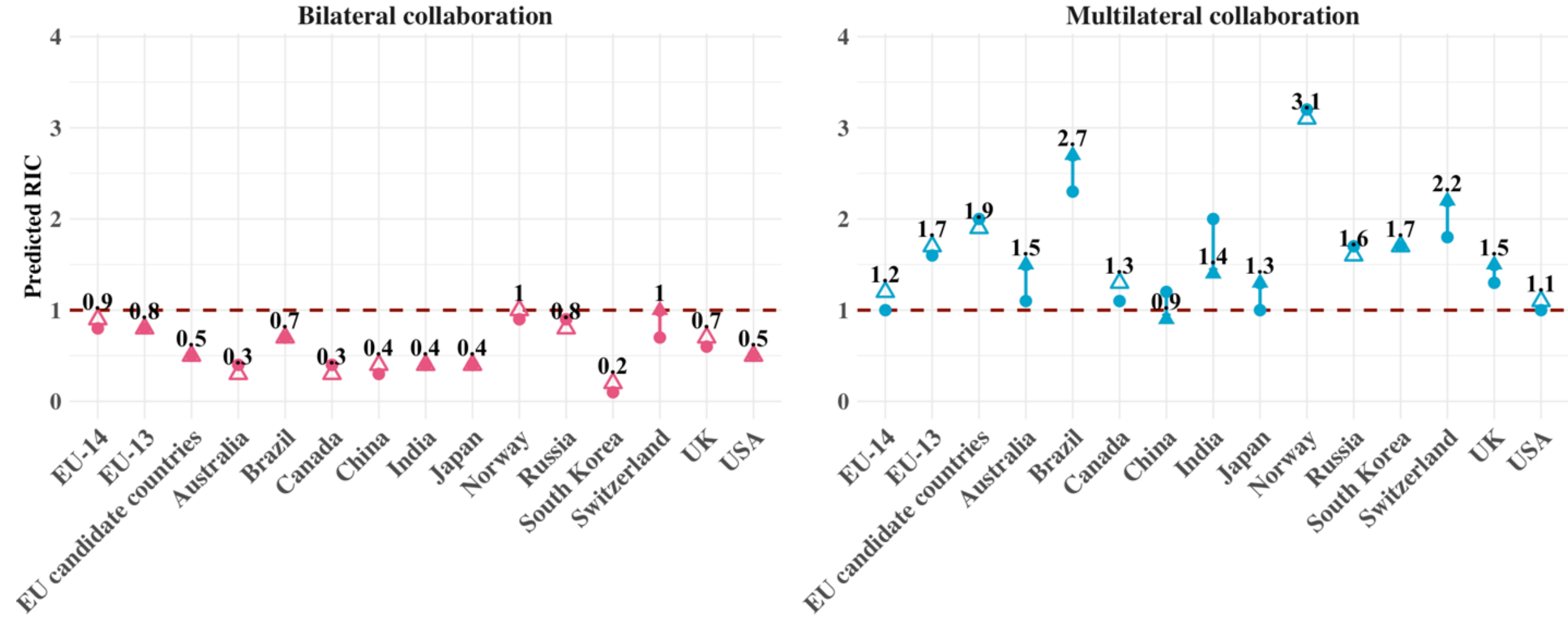

Physics & astronomy
Bilateral collaboration
Multilateral collaboration
Predicted RIC
0
1
2
3
4
0.9
0.8
0.5
0.3
0.7
0.3
0.4
0.4
0.4
1
0.8
0.2
1
0.7
0.5
1.2
1.7
1.9
1.5
2.7
1.3
0.9
1.4
1.3
3.1
1.6
1.7
2.2
1.5
1.1
EU-14
EU-13
EU candidate countries
Australia
Brazil
Canada
China
India
Japan
Norway
Russia
South Korea
Switzerland
UK
USA

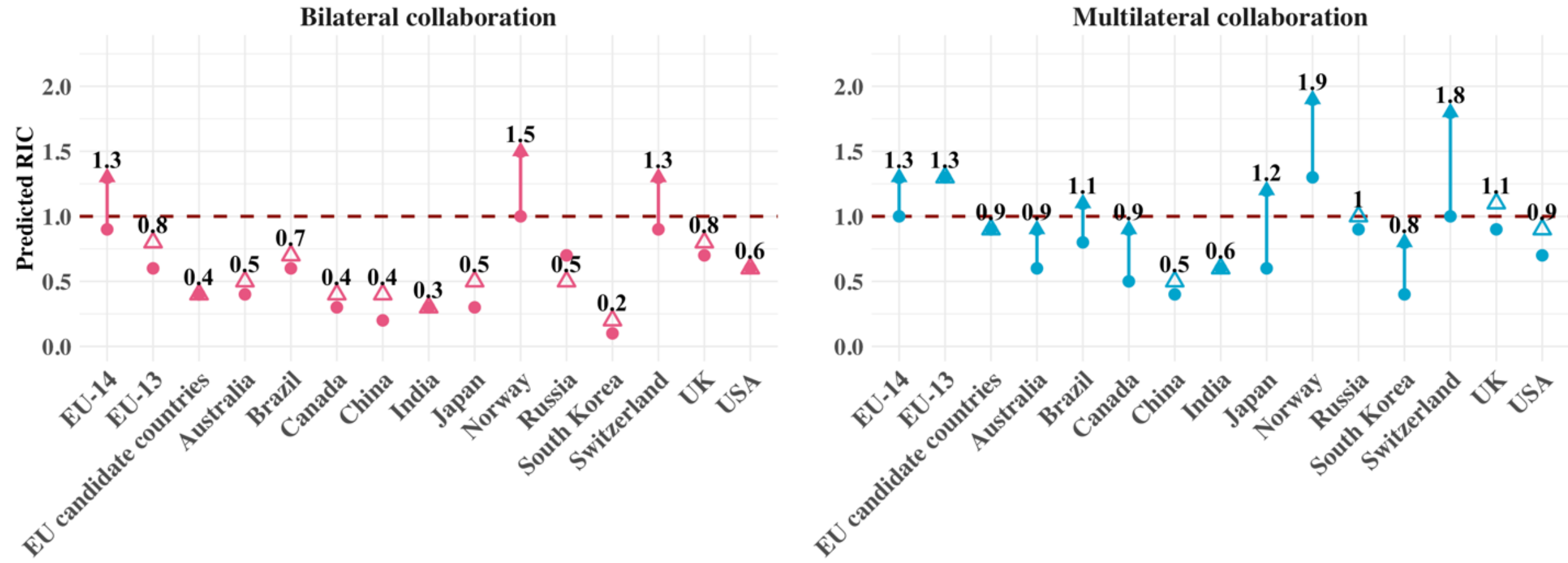

**Fig. 6.** Predicted RIC, 2010-2024.

A Model 2 linear mixed-effects model, with country as a random intercept was fitted to estimate predicted RIC values in 2010 and 2024. Similar to Model 1, Model 2 explained nearly all variance in RIC values ($R^2$_m = 0.972), while the inclusion of country-level random effects increased the explained variance only marginally ($R^2$_c = 0.973), indicating that temporal, disciplinary, and partner-specific factors accounted for most observed variation.

At the aggregate level, RIC increased for both bilateral and multilateral collaboration between 2010 and 2024 (Fig. 6). However, the increase was substantially stronger for multilateral collaboration (0.94 → 1.220, Δ = +0.28) than for bilateral collaboration (0.57 → 0.68, Δ = +0.11).

Disciplinary differences in RIC evolution were pronounced. For bilateral collaboration, the strongest increases were observed in social sciences & humanities (+0.15), medicine (+0.14), and computer science (+0.14), while physics & astronomy showed only a small, and statistically non-significant increase. Despite these increases, bilateral RIC remained below 1.0 across all disciplines, indicating that bilateral collaboration generally remained less intensive than expected.

By contrast, multilateral collaboration increased across all disciplines, with RIC values exceeding 1.0 in several fields in 2024. The strongest growth was observed in medicine (Δ = +0.40), followed by life & environmental sciences (Δ = +0.32) and engineering (Δ = +0.30). Despite more modest growth (1.59 → 1.67), physics & astronomy maintained the highest multilateral RIC, reflecting the highly internationalised and consortium-based nature of research in this field.

For bilateral collaboration, the strongest increases were concentrated within Europe, particularly within EU-14 countries and with Switzerland and Norway, which maintained the highest RIC values across all disciplines. These increases were especially pronounced in computer science, where bilateral RIC within the EU-14 countries rose from 0.94 to 1.53 and with Switzerland from 0.82 to 1.35.

For multilateral collaboration, the largest increases were observed in engineering and computer science with Switzerland and Norway, as well as in medicine and social sciences & humanities with Japan and South Korea. In medicine, multilateral RIC with Japan increased from 0.54 to 1.32, while collaboration with South Korea rose from 0.43 to 1.12. In contrast to the general upward trend, several declines were observed in physics & astronomy, particularly in collaboration with China (1.16 → 0.93) and India (2.03 → 1.36).

Overall, the results demonstrate that the evolution of collaboration intensity between 2010 and 2024 was characterised not only by increasing multilateral collaboration, but also by substantial disciplinary and geopolitical differentiation.

The top 10 increases and decreases in RIC across collaboration partners and disciplines are reported in the Appendix (Tables 4 and 5).

### 4.5. Empirical RIC in 2024

Figure 7 illustrates empirical RIC values in 2024. For bilateral collaboration, most partner countries exhibited RIC below the expected level (< 1.0). However, within the EU-14, RIC either reached or exceeded the expected level across all disciplines except physics & astronomy, corresponding to comparatively high collaboration rates within the EU-14 (~30-40%) across disciplines. Slightly lower RIC values, though still close to the expected level, were observed for collaboration with the EU-13, whereas substantially lower values were observed for EU candidate countries.

While bilateral collaboration rates with the USA slightly exceeded 10%, the RIC was below the expected level (~0.4-0.6), with the highest RIC values observed for Germany, the Netherlands, and Italy. Similar patterns were for collaboration with the UK (~4-10%; RIC = ~0.4-0.8) and China (~5-10% RIC=~0.3-0.7). In contrast, Switzerland and Norway exhibited relatively low collaboration rates (~2–3% and ~1%, respectively) but RIC values above the expected level. The lowest bilateral RIC values were observed with South Korea (~0.1-0.2).

Overall, multilateral RIC values exceeded bilateral ones across all disciplines and collaboration partners, consistent with the patterns for collaboration rates. Disciplinary differences in multilateral RIC closely mirrored those for collaboration rates. The highest RIC values were found in physics & astronomy and medicine, followed by life & environmental sciences and social sciences & humanities, whereas computer science and engineering displayed comparatively lower levels of multilateral collaboration intensity.

Within the EU-14, multilateral RIC values generally exceeded the expected level but remained below 2.0 across disciplines. Collaboration with the EU-13 exhibited even higher intensity in some disciplines, particularly medicine, where RIC approached or exceeded 2.0. By contrast, collaboration with EU candidate countries remained below the expected level in most disciplines, especially in computer science and engineering, while physics & astronomy again represented an exception with RIC values above 2.0.

For multilateral collaboration with the USA and the UK, most RIC values either reached or exceeded the expected level, with the highest values observed in physics & astronomy and the lowest in social sciences & humanities. Overall, RIC values were relatively similar across the nine EU-14 countries. Even higher RIC values were with Switzerland, particularly in collaboration with Austria and Germany, which also exhibited the highest collaboration rates. A similar pattern was observed with Norway, where Finland and Sweden showed the highest RIC values.

In contrast, multilateral RIC with China reached the expected level only in physics & astronomy, while remaining below this level in all other disciplines. A similar pattern was observed for South Korea and India. With Australia and Canada, RIC values were generally close to the expected level, occasionally exceeding it, depending on the discipline and partner country.

Multilateral RIC values with Brazil exceeded the expected level across all disciplines. Nevertheless, collaboration rates with Brazil generally remained below 10%, except in physics & astronomy. This suggests that even relatively infrequent collaboration can represent a strong collaborative preference when it occurs more often than expected given the overall structure of international collaboration.

## Computer science

### Bilateral collaboration

| | Austria | Belgium | Finland | France | Germany | Italy | Netherlands | Spain | Sweden |
|---|---|---|---|---|---|---|---|---|---|
| EU-14 | 2.2 | 1.9 | 1.4 | 1.1 | 1.3 | 1.6 | 1.7 | 1.4 | 1.4 |
| EU-13 | 1.6 | 0.7 | 1.1 | 1.1 | 1 | 1 | 0.8 | 1 | 0.6 |
| EU candidate countries | 0.9 | 0.4 | 1.2 | 0.5 | 0.6 | 0.7 | 0.7 | 1 | 0.9 |
| Australia | 0.3 | 0.2 | 0.3 | 0.2 | 0.3 | 0.2 | 0.5 | 0.3 | 0.3 |
| Brazil | 0.4 | 1.5 | 0.4 | 1.6 | 0.6 | 1 | 0.9 | 1.4 | 1.1 |
| Canada | 0.3 | 0.3 | 0.2 | 0.7 | 0.5 | 0.3 | 0.4 | 0.2 | 0.6 |
| China | 0.2 | 0.4 | 0.7 | 0.4 | 0.4 | 0.4 | 0.5 | 0.3 | 0.7 |
| India | 0.4 | 0.4 | 0.9 | 0.6 | 0.4 | 0.6 | 0.3 | 0.4 | 0.7 |
| Japan | 0.7 | 0.3 | 0.3 | 0.8 | 0.6 | 0.3 | 0.4 | 0.2 | 0.3 |
| Norway | 0.9 | 0.8 | 2.8 | 1 | 1 | 0.9 | 1 | 1 | 2.3 |
| Russia | 1.2 | 0.2 | 0.2 | 0.7 | 0.5 | 0.3 | 0.3 | 0.5 | 0.4 |
| South Korea | 0.1 | 0.2 | 0.4 | 0.2 | 0.3 | 0.2 | 0.1 | 0.2 | 0.4 |
| Switzerland | 1.4 | 1.1 | 0.4 | 1.6 | 2.6 | 1.8 | 1.3 | 0.7 | 0.5 |
| UK | 0.4 | 0.8 | 0.6 | 0.5 | 0.9 | 0.9 | 0.8 | 0.6 | 0.7 |
| USA | 0.6 | 0.4 | 0.5 | 0.6 | 0.8 | 0.6 | 0.6 | 0.4 | 0.4 |

### Multilateral collaboration

| | Austria | Belgium | Finland | France | Germany | Italy | Netherlands | Spain | Sweden |
|---|---|---|---|---|---|---|---|---|---|
| EU-14 | 1.9 | 1.8 | 1.3 | 1.5 | 1.8 | 1.7 | 1.8 | 1.5 | 1.4 |
| EU-13 | 2.1 | 1.6 | 1.3 | 1.2 | 1.2 | 1.4 | 1.1 | 1.6 | 1 |
| EU candidate countries | 1.1 | 1 | 1.1 | 0.7 | 0.6 | 0.9 | 0.6 | 1.1 | 0.9 |
| Australia | 0.6 | 0.8 | 0.8 | 0.6 | 0.7 | 0.7 | 1 | 0.7 | 1 |
| Brazil | 1.9 | 1.4 | 1.4 | 1.8 | 1.4 | 1.7 | 1.6 | 2.4 | 1.2 |
| Canada | 0.7 | 1 | 0.9 | 0.9 | 0.9 | 0.8 | 1 | 0.7 | 0.7 |
| China | 0.3 | 0.4 | 0.6 | 0.5 | 0.5 | 0.4 | 0.4 | 0.5 | 0.5 |
| India | 0.5 | 0.4 | 0.6 | 0.5 | 0.4 | 0.7 | 0.4 | 0.6 | 0.7 |
| Japan | 1 | 1 | 0.8 | 1 | 1 | 1 | 1.1 | 1 | 0.8 |
| Norway | 2 | 2 | 2.8 | 1.2 | 1.6 | 1.6 | 2.6 | 1.4 | 3.4 |
| Russia | 0.8 | 1.5 | 0.8 | 1 | 0.9 | 1 | 0.5 | 0.7 | 0.9 |
| South Korea | 0.6 | 0.5 | 0.9 | 0.5 | 0.4 | 0.5 | 0.3 | 0.9 | 0.6 |
| Switzerland | 2.8 | 2.8 | 1.4 | 2.1 | 3.5 | 2.4 | 2.5 | 1.8 | 1.7 |
| UK | 1.1 | 1.2 | 1.1 | 1 | 1.2 | 1.1 | 1.4 | 1 | 1.2 |
| USA | 1 | 1 | 1 | 0.9 | 1.1 | 1 | 1.1 | 0.9 | 1 |

## Engineering

### Bilateral collaboration

| | Austria | Belgium | Finland | France | Germany | Italy | Netherlands | Spain | Sweden |
|---|---|---|---|---|---|---|---|---|---|
| EU-14 | 1.9 | 1.7 | 1.2 | 1 | 1.1 | 1.5 | 1.5 | 1.5 | 1.2 |
| EU-13 | 2 | 0.9 | 1.3 | 0.8 | 1 | 1.3 | 0.5 | 0.8 | 0.8 |
| EU candidate countries | 1.2 | 0.3 | 1.4 | 0.4 | 0.8 | 0.9 | 0.5 | 0.4 | 0.5 |
| Australia | 0.2 | 0.2 | 0.3 | 0.2 | 0.3 | 0.3 | 0.3 | 0.2 | 0.2 |
| Brazil | 0.5 | 0.8 | 0.7 | 1.5 | 0.9 | 0.9 | 0.8 | 2.2 | 1.1 |
| Canada | 0.1 | 0.4 | 0.4 | 0.8 | 0.5 | 0.4 | 0.3 | 0.2 | 0.3 |
| China | 0.3 | 0.7 | 0.9 | 0.6 | 0.7 | 0.4 | 0.7 | 0.3 | 0.9 |
| India | 0.3 | 0.4 | 0.8 | 0.4 | 0.7 | 0.5 | 0.5 | 0.4 | 0.8 |
| Japan | 0.5 | 0.3 | 0.7 | 0.8 | 0.6 | 0.3 | 0.3 | 0.3 | 0.3 |
| Norway | 1.4 | 0.4 | 1.3 | 0.5 | 1.1 | 0.9 | 1.3 | 0.8 | 2.2 |
| Russia | 0.9 | 0.4 | 0.2 | 1 | 1 | 0.6 | 0.4 | 0.4 | 0.3 |
| South Korea | 0.1 | 0.3 | 0.3 | 0.2 | 0.3 | 0.1 | 0.1 | 0.1 | 0.3 |
| Switzerland | 1.4 | 1.1 | 0.4 | 1.3 | 2.3 | 2 | 1.4 | 0.9 | 0.8 |
| UK | 0.5 | 0.5 | 0.6 | 0.4 | 0.6 | 0.8 | 0.8 | 0.6 | 0.7 |
| USA | 0.4 | 0.4 | 0.3 | 0.5 | 0.7 | 0.6 | 0.7 | 0.4 | 0.4 |

### Multilateral collaboration

| | Austria | Belgium | Finland | France | Germany | Italy | Netherlands | Spain | Sweden |
|---|---|---|---|---|---|---|---|---|---|
| EU-14 | 1.7 | 1.7 | 1.3 | 1.5 | 1.6 | 1.6 | 1.7 | 1.6 | 1.5 |
| EU-13 | 1.7 | 1.1 | 1.5 | 1.2 | 1.3 | 1.4 | 1 | 1.2 | 1.2 |
| EU candidate countries | 1 | 0.6 | 0.8 | 0.6 | 0.7 | 0.8 | 0.5 | 0.8 | 0.7 |
| Australia | 0.8 | 0.8 | 0.8 | 0.7 | 0.8 | 0.6 | 1 | 0.7 | 0.9 |
| Brazil | 1.3 | 1.4 | 1.6 | 1.2 | 1 | 1.5 | 1.2 | 2 | 1.2 |
| Canada | 1 | 0.8 | 1.2 | 0.9 | 0.9 | 0.9 | 1.1 | 0.7 | 0.9 |
| China | 0.4 | 0.6 | 0.7 | 0.6 | 0.7 | 0.5 | 0.6 | 0.5 | 0.7 |
| India | 0.4 | 0.2 | 0.4 | 0.3 | 0.4 | 0.5 | 0.3 | 0.4 | 0.5 |
| Japan | 1 | 0.8 | 1.2 | 1.1 | 1.3 | 0.9 | 0.9 | 0.9 | 1 |
| Norway | 2.5 | 1.2 | 3.1 | 1.1 | 1.5 | 1.4 | 2.1 | 1.3 | 3 |
| Russia | 0.9 | 0.9 | 0.7 | 0.7 | 0.7 | 0.7 | 0.4 | 0.9 | 0.7 |
| South Korea | 0.4 | 0.5 | 0.5 | 0.4 | 0.4 | 0.4 | 0.4 | 0.4 | 0.5 |
| Switzerland | 3 | 2.3 | 2 | 2.4 | 3.5 | 2.4 | 2.6 | 1.9 | 2.1 |
| UK | 1.1 | 1.3 | 1.2 | 1.2 | 1.3 | 1.2 | 1.4 | 1.2 | 1.1 |
| USA | 1 | 0.9 | 0.9 | 1 | 1.1 | 1 | 1.2 | 0.9 | 0.9 |

## Life & environmental sciences

### Bilateral collaboration

| | Austria | Belgium | Finland | France | Germany | Italy | Netherlands | Spain | Sweden |
|---|---|---|---|---|---|---|---|---|---|
| EU-14 | 1.6 | 1.4 | 1.1 | 1 | 1 | 1.3 | 1.3 | 1.3 | 1.2 |
| EU-13 | 1.9 | 0.8 | 1.4 | 0.6 | 1 | 1.4 | 0.5 | 0.8 | 0.8 |
| EU candidate countries | 1 | 0.2 | 1.1 | 0.3 | 0.8 | 1.1 | 0.4 | 0.5 | 0.6 |
| Australia | 0.3 | 0.3 | 0.3 | 0.4 | 0.4 | 0.3 | 0.4 | 0.3 | 0.4 |
| Brazil | 0.2 | 0.5 | 0.4 | 0.9 | 0.6 | 0.6 | 0.5 | 1.5 | 0.6 |
| Canada | 0.2 | 0.3 | 0.3 | 0.8 | 0.3 | 0.3 | 0.4 | 0.2 | 0.3 |
| China | 0.2 | 0.6 | 0.8 | 0.4 | 0.5 | 0.2 | 0.6 | 0.2 | 0.7 |
| India | 0.2 | 0.2 | 0.6 | 0.4 | 0.6 | 0.4 | 0.3 | 0.2 | 0.6 |
| Japan | 0.4 | 0.3 | 0.4 | 0.6 | 0.4 | 0.3 | 0.3 | 0.3 | 0.4 |
| Norway | 0.6 | 0.5 | 1.8 | 0.7 | 0.9 | 0.6 | 1.1 | 0.7 | 3.5 |
| Russia | 0.7 | 0.9 | 0.9 | 0.9 | 0.7 | 0.5 | 0.4 | 0.2 | 0.7 |
| South Korea | 0.3 | 0.3 | 0.1 | 0.3 | 0.3 | 0.1 | 0.1 | 0.1 | 0.2 |
| Switzerland | 1.5 | 0.9 | 0.8 | 1.5 | 2.2 | 1.4 | 1 | 0.7 | 0.5 |
| UK | 0.5 | 0.5 | 0.6 | 0.6 | 0.7 | 0.9 | 0.8 | 0.8 | 0.8 |
| USA | 0.5 | 0.4 | 0.4 | 0.5 | 0.6 | 0.6 | 0.6 | 0.4 | 0.5 |

### Multilateral collaboration

| | Austria | Belgium | Finland | France | Germany | Italy | Netherlands | Spain | Sweden |
|---|---|---|---|---|---|---|---|---|---|
| EU-14 | 1.4 | 1.4 | 1.2 | 1.3 | 1.3 | 1.4 | 1.4 | 1.4 | 1.2 |
| EU-13 | 2.1 | 1.6 | 2 | 1.3 | 1.5 | 1.7 | 1.2 | 1.5 | 1.4 |
| EU candidate countries | 1.3 | 1 | 0.9 | 0.8 | 0.9 | 1.5 | 0.7 | 1.1 | 0.8 |
| Australia | 0.9 | 1 | 1.1 | 1 | 0.9 | 0.9 | 1.1 | 1.1 | 1.1 |
| Brazil | 1 | 1.1 | 1 | 1.3 | 1.1 | 1.2 | 1.2 | 1.8 | 1.1 |
| Canada | 1.1 | 1.2 | 1.2 | 1.4 | 1.1 | 1.1 | 1.3 | 1.1 | 1.2 |
| China | 0.5 | 0.6 | 0.7 | 0.6 | 0.6 | 0.5 | 0.6 | 0.5 | 0.7 |
| India | 0.5 | 0.5 | 0.6 | 0.5 | 0.5 | 0.6 | 0.4 | 0.4 | 0.6 |
| Japan | 1.3 | 1.2 | 1.2 | 1.3 | 1.1 | 1.1 | 1 | 1 | 0.9 |
| Norway | 2 | 1.7 | 4.2 | 1.7 | 1.8 | 1.5 | 2 | 1.7 | 3.7 |
| Russia | 1.3 | 1.1 | 1.7 | 1 | 1.2 | 1.1 | 0.8 | 1 | 1.2 |
| South Korea | 0.8 | 0.8 | 0.7 | 0.7 | 0.7 | 0.8 | 0.5 | 0.8 | 0.6 |
| Switzerland | 2.5 | 2.1 | 1.9 | 2 | 2.3 | 1.9 | 1.9 | 1.6 | 1.8 |
| UK | 1.2 | 1.3 | 1.3 | 1.3 | 1.2 | 1.3 | 1.5 | 1.3 | 1.3 |
| USA | 1 | 1 | 0.9 | 1 | 1.1 | 1 | 1.1 | 1 | 1 |

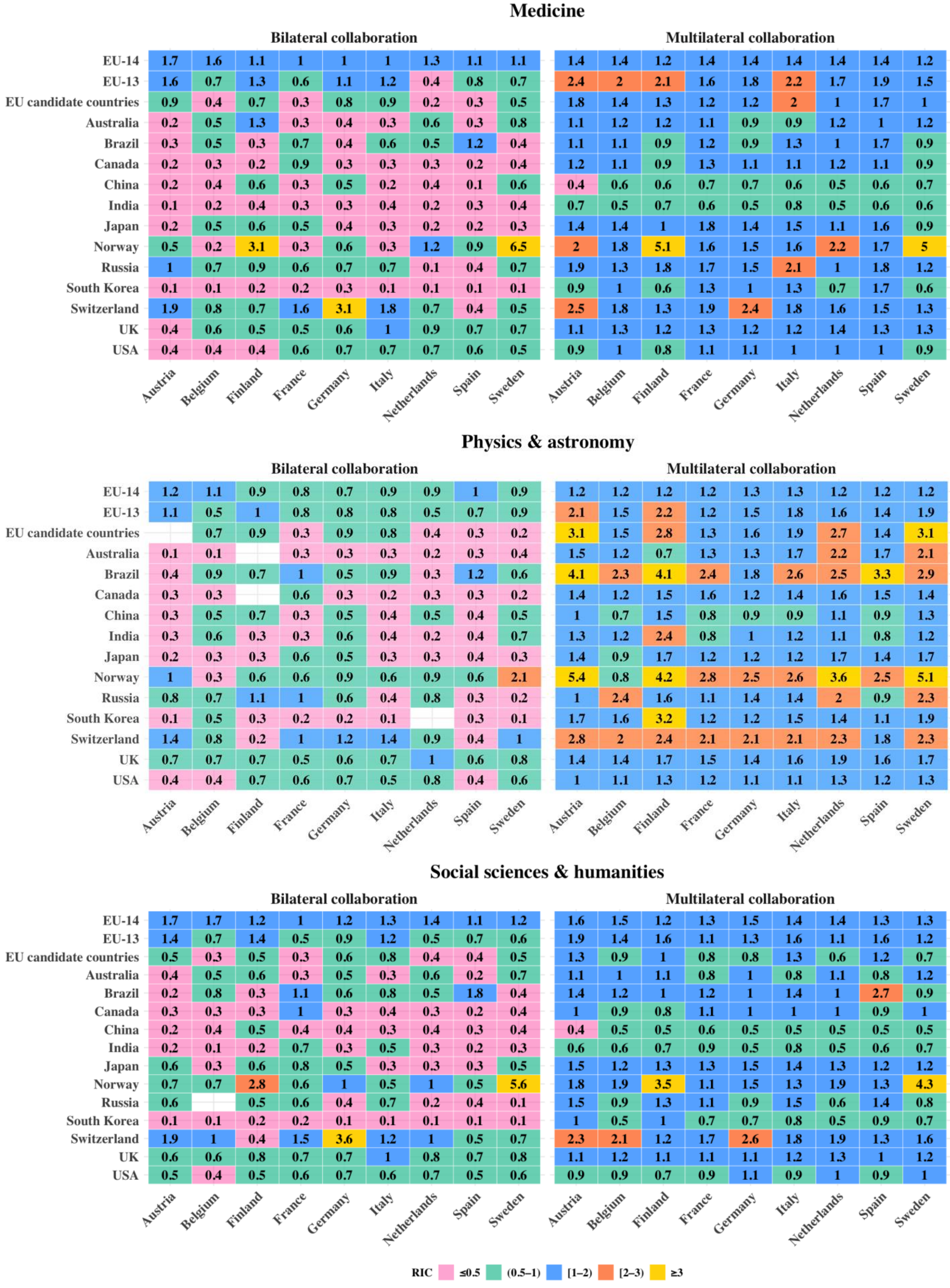

**Fig. 7.** RIC in 2024.

## 4.6 Relationship between changes in collaboration rate and RIC

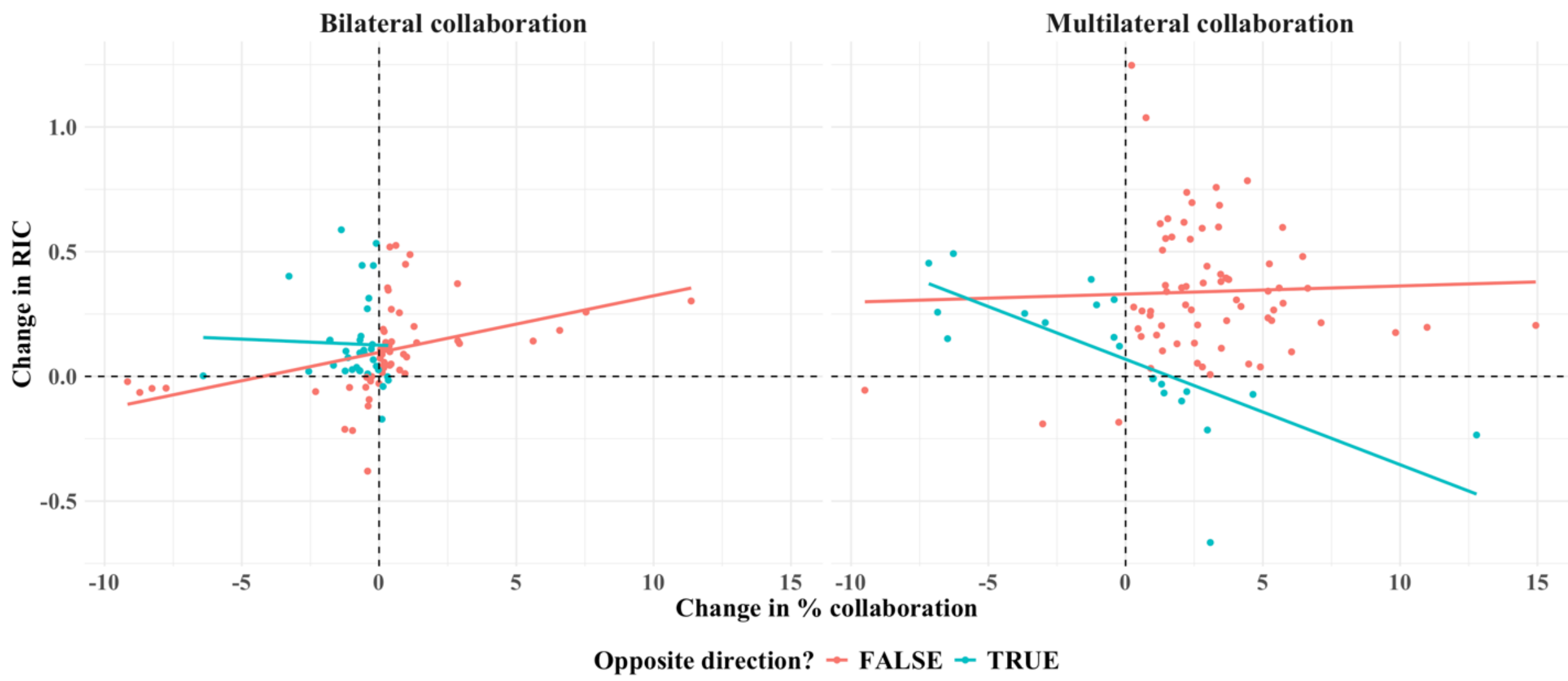

**Fig. 8.** Relationship between changes in collaboration rate (%) and RIC, by collaboration type, 2010–2024.

The comparison between changes in collaboration rate and RIC (Fig. 8) revealed distinct patterns for bilateral and multilateral collaboration. For bilateral collaboration, the association between Δ% collaboration and ΔRIC was generally positive. The fitted regression line intersected the x-axis at approximately –4.2%, indicating that moderate declines in collaboration rate were not necessarily associated with a decrease in RIC. Increases in collaboration rate were typically associated with small increases in RIC.

Cases showing opposite-direction changes were concentrated among declining collaboration rates. For the EU-14, the UK, the USA, Japan, and Canada, RIC increased despite declining collaboration rates. By contrast, Russia and India showed the opposite pattern, with increasing collaboration rates accompanied by declining RIC values.

For multilateral collaboration, ΔRIC also increased across most of the observed range of Δ% collaboration, although changes were generally smaller. When Δ% collaboration remained close to zero, RIC often still increased slightly, indicating that RIC could strengthen even without substantial growth in collaboration rate. Similar to bilateral collaboration, RIC increased despite declining collaboration rates for the EU-14, the USA, Japan, and Canada. In contrast, EU candidate countries, South Korea, India, and China showed increasing collaboration rates combined with declining RIC, particularly in physics & astronomy.
Overall, increases in collaboration rate were generally associated with increases in RIC across both collaboration types, although the relationship was stronger and more stable for bilateral collaboration.

### 4.7.Geopolitical disruptions and international collaboration: Brexit and Russia’s full-scale invasion of Ukraine

Across all disciplines, the predicted collaboration rate and RIC between the nine EU-14 countries and Russia remained largely stable between 2021 and 2024, particularly for bilateral collaboration (Figs. 9–10). The main changes were observed in multilateral collaboration, where the collaboration rate in physics & astronomy declined from 16.9% to 14.8%, while modest increases were observed in medicine (from 4.2% to 4.7%) and social sciences & humanities (from 2.9% to 3.2%). Correspondingly, multilateral RIC increased in medicine (from 1.48 to 1.63) and social sciences & humanities (from 1.01 to 1.04), while remaining stable in physics & astronomy. Confidence intervals for both collaboration rate and RIC are relatively wide, indicating substantial cross-country variation.

Figures 11 and 12 show changes in collaboration rate and RIC between the nine EU-14 countries and the UK. Across disciplines, both indicators remained highly stable between 2018 and 2024. Multilateral collaboration consistently exceeded bilateral collaboration and showed only marginal increases over time, whereas bilateral collaboration rates exhibited only minor declines.

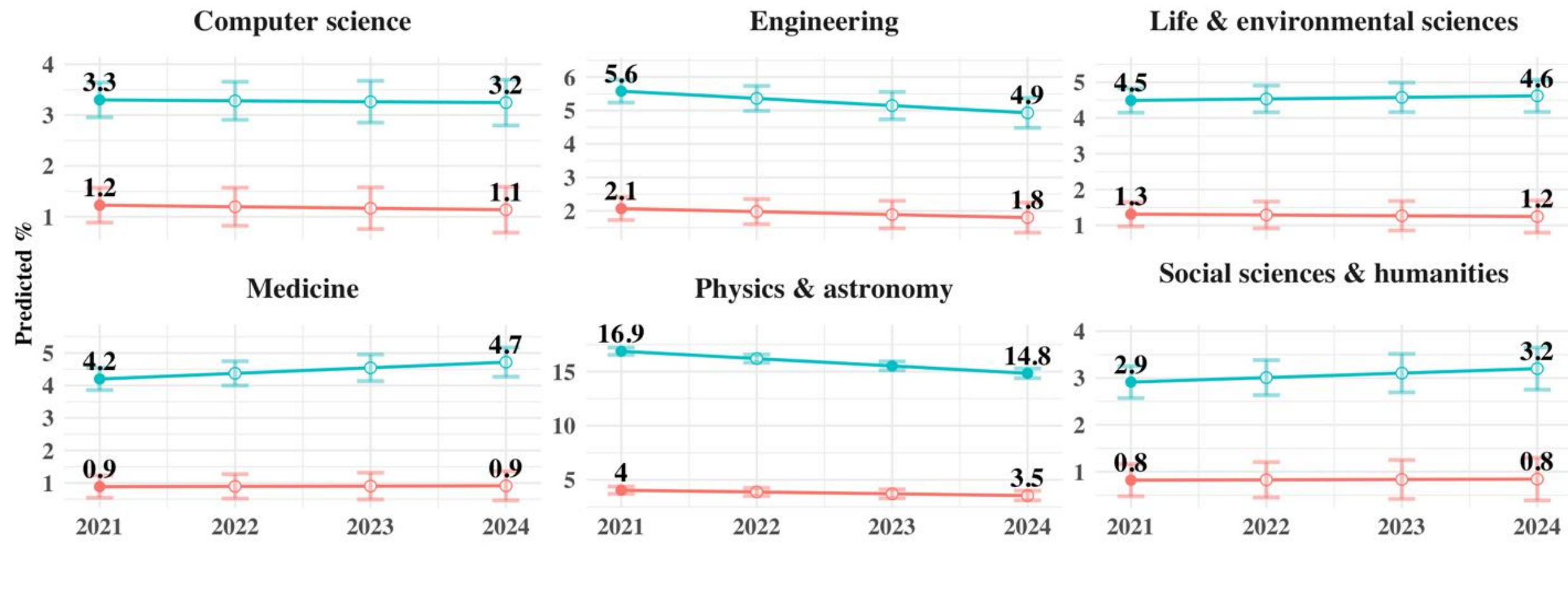

**Fig. 9.** Predicted collaboration rate with Russia, 2021-2024.

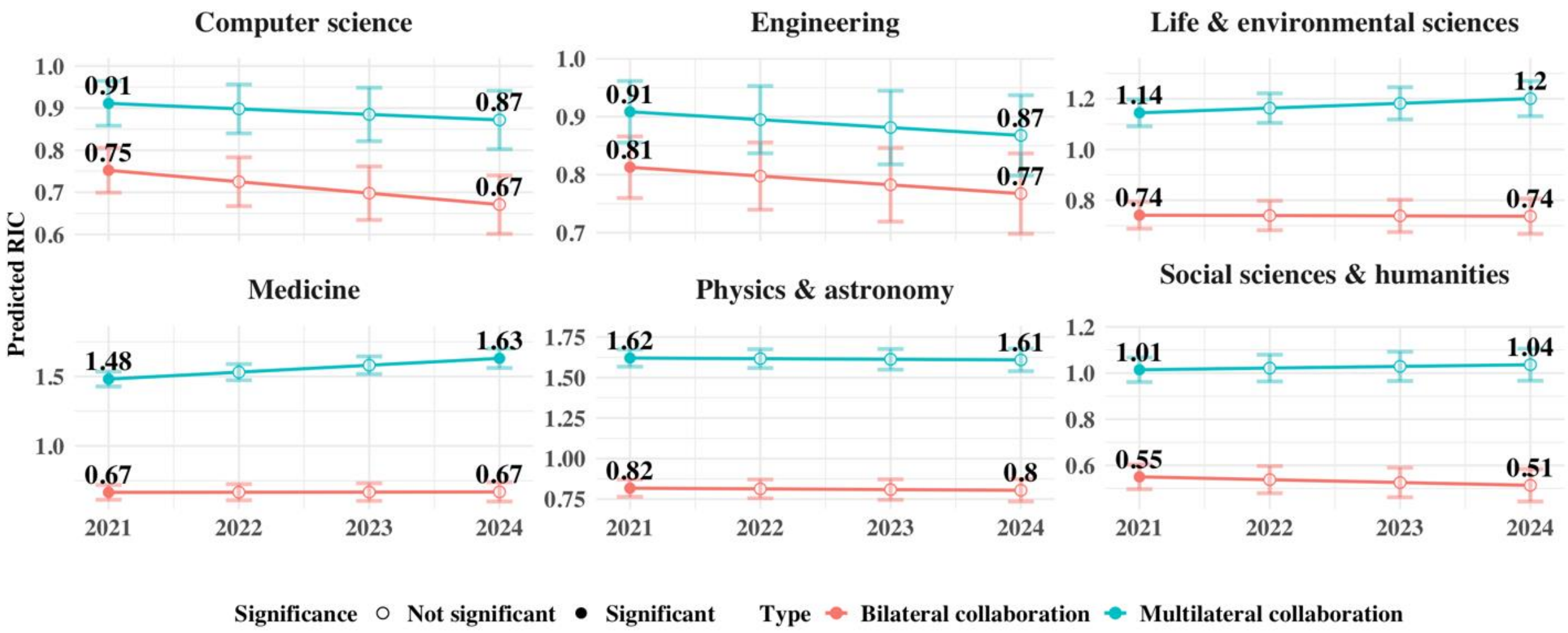

**Fig. 10.** Predicted RIC with Russia, 2021-2024.

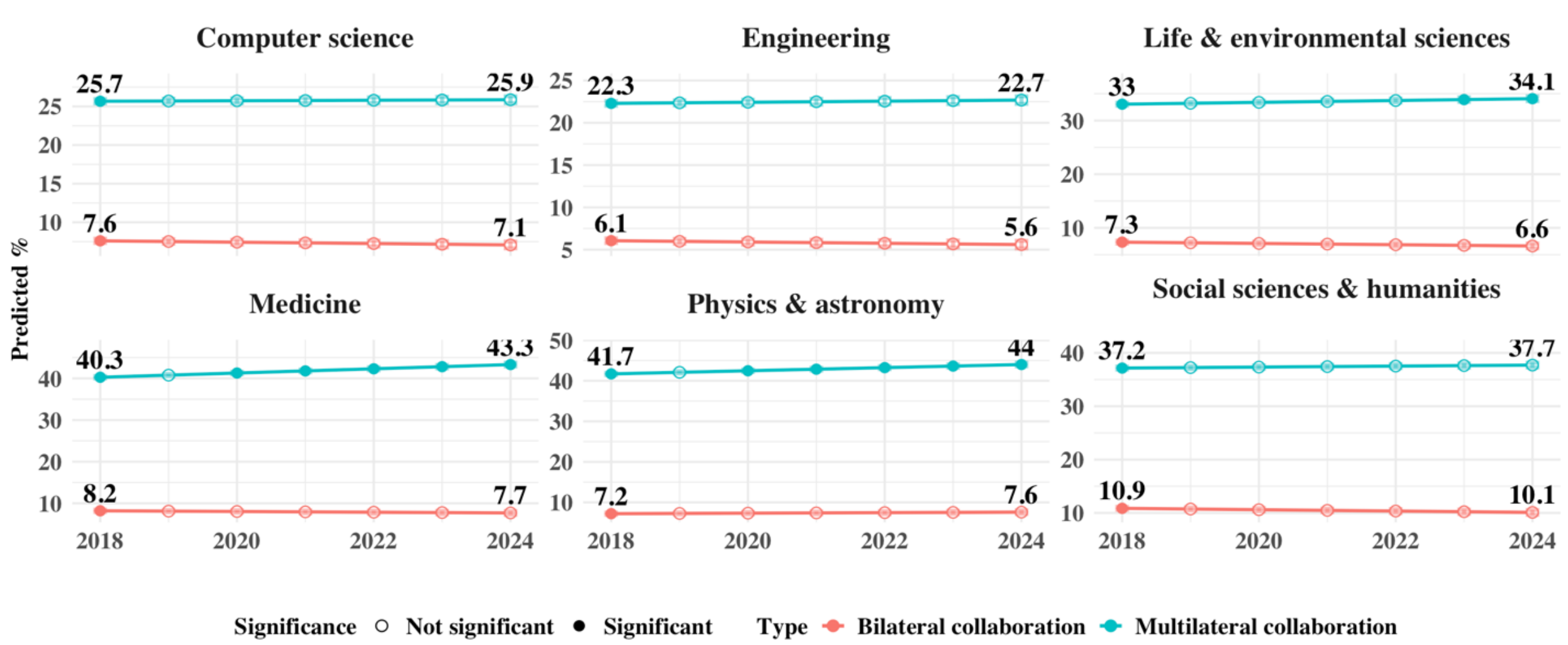

**Fig. 11**. Collaboration rate with the UK, 2021-2024.

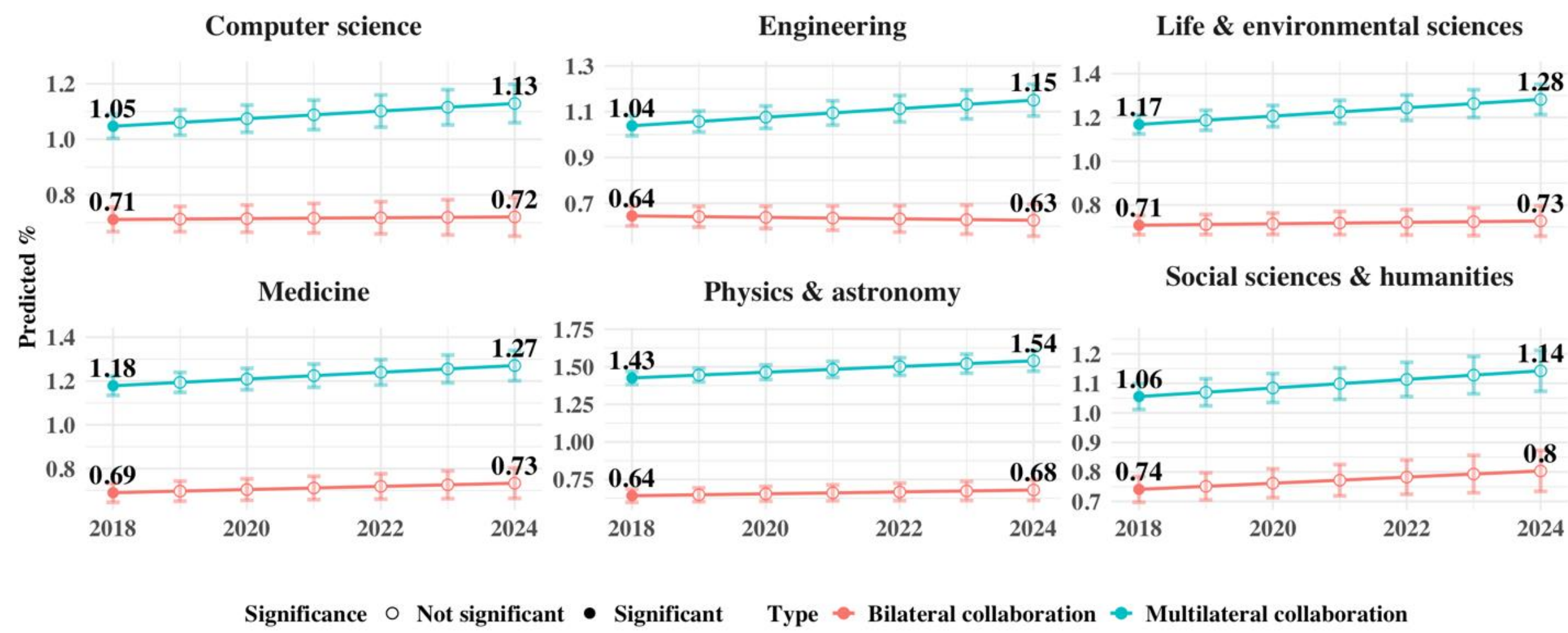

**Fig. 12.** RIC with the UK, 2021-2024.

## 5. DISCUSSION AND CONCLUSIONS

This study provides a comprehensive analysis of scientific collaboration of nine EU-14 countries across six disciplines between 2010 and 2024. By examining collaboration rate and RIC across disciplines and global partners, several key patterns emerged.

Bilateral collaboration rates remained relatively stable and predominantly concentrated within EU-14 countries, followed by the USA, the UK, and China. In contrast, multilateral collaboration rates increased significantly across all disciplines, with the highest increase observed in medicine and the highest overall rates maintained in physics & astronomy. This trend reflects the growing importance of large-scale international research consortia in infrastructure-intensive fields that address global scientific and societal challenges (Hsiehchen et al., 2015; Adams & Szomszor, 2022; Lima-Toivanen et al., 2025). The strongest growth in both bilateral and multilateral collaboration rates was observed with China, particularly in engineering, life & environmental sciences, and computer science, disciplines that closely align with areas prioritised in China's science and innovation policies (Lu, 2012)

RIC increased for both bilateral and multilateral collaboration between 2010 and 2024. For bilateral collaboration, the highest RIC was observed within EU-14 countries, as well as with Norway and Switzerland, highlighting the role of cultural, socio-economic, and geographic proximity in shaping international collaboration patterns. The positive influence of these proximity factors was evident in the relatively high collaboration intensity between Spain and Brazil; Sweden, Finland and Norway; Austria, Germany, Italy and Switzerland.

RIC growth was stronger for multilateral collaboration, consistent with Okamura's (2023) "shrinking world" or "global village" of Yao et al. (2019), which can be attributed to the effects of Horizon 2020 and Horizon Europe in promoting international scientific collaboration (Barre et al. 2013; Kwiek 2021). Similar to the trends observed for collaboration rates, medicine experienced the largest increase in multilateral RIC, while the highest RIC remained in physics & astronomy. The lowest RIC was observed with South Korea, India, and China, countries characterised by weaker socio-economic, geographic, and language proximity to the EU-14 countries.

In terms of European integration, collaboration patterns revealed a clear hierarchical structure, with the highest RIC observed within the EU-14, intermediate levels with the EU-13 countries, and the lowest levels with EU candidate countries. While both collaboration rates and RIC increased substantially for the EU-13, the corresponding increases for EU candidate countries were comparatively modest. Based on collaboration rates, the integration of EU-13 and EU candidate countries into the European research system occurred primarily through multilateral collaboration rather than bilateral partnerships. Overall, the findings suggest that European scientific integration has progressed unevenly, with EU-13 countries becoming increasingly engaged with the nine EU-14 countries included in this study, while EU candidate countries continue to occupy a more peripheral position within the European research landscape.

No substantial changes in collaboration rates or RIC with the UK were observed following Brexit during the analysed period, which is consistent with the findings of Hladchenko (2026b). This stability may be attributed to the UK's continued association with Horizon 2020 and Horizon Europe. By contrast, a decline in multilateral collaboration rates in physics & astronomy with Russia coincided temporally with Russia's suspension from Horizon Europe. However, collaboration rates with Russia increased in medicine, while RIC across disciplines remained close to or above the expected level. One possible interpretation is that, despite the suspension of Russian organisations from Horizon Europe projects, individual researchers continued to engage in international scientific collaboration.

Across both collaboration types, increases in collaboration rates were generally accompanied by increases in RIC, while declines in collaboration rates were often associated with reductions in RIC. Some established collaboration partners maintained relatively high RIC despite fluctuations in collaboration rates, whereas some emerging partners increased collaboration rates without corresponding increases in RIC.

Overall, international collaboration did not evolve uniformly. Growth was driven disproportionately by multilateral partnerships, leading scientific countries, and disciplines such as medicine and physics & astronomy. These findings suggest increasing importance for countries to participate in large-scale international research networks. Policies that facilitate access to multilateral research infrastructures, international consortia, and cross-disciplinary collaboration, therefore, play a critical role in shaping the future of the global research landscape.

## Appendix

Table 1 Mapping of OpenAlex research fields to aggregated disciplinary categories

| Discipline | Research subfield, OpenAlex classification |
| --- | --- |
| Computer science | Computer science |
| Engineering | Chemical engineering |
| | Energy |
| | Engineering |
| | Materials Science |
| | Agricultural & biological sciences |
| | Biochemistry, genetics and molecular biology |

| Life & environmental sciences | Environmental sciences |
|---|---|
| Medicine | Medicine |
| | Dentistry |
| Physics & astronomy | Physics & astronomy |
| Social sciences & humanities | Business, management and accounting |
| | Economics, Econometrics and finance |
| | Psychology |
| | Social sciences |
| | Arts & humanities |

Table 2. Top 10 increases in collaboration rate, 2010-2024

| Type | Discipline | Country | 2010 (%) | 2024 (%) | Δ (pp) | Δ (%) | 95% CI lower | 95% CI upper | p_value | significant |
|---|---|---|---|---|---|---|---|---|---|---|
| Multilateral collaboration | Engineering | China | 13,5 | 28,4 | 14,9 | 110,7 | 13,5 | 16,3 | < 0.001 | TRUE |
| Multilateral collaboration | Physics & astronomy | China | 19,5 | 32,3 | 12,8 | 65,5 | 11,4 | 14,2 | < 0.001 | TRUE |
| Bilateral collaboration | Engineering | China | 9,1 | 20,5 | 11,4 | 125,1 | 10,0 | 12,8 | < 0.001 | TRUE |
| Multilateral collaboration | Computer science | China | 11,2 | 22,1 | 11,0 | 98,3 | 9,6 | 12,4 | < 0.001 | TRUE |
| Multilateral collaboration | Life & environmental sciences | China | 9,5 | 19,4 | 9,8 | 103,2 | 8,4 | 11,2 | < 0.001 | TRUE |
| Bilateral collaboration | Life & environmental sciences | China | 5,8 | 13,4 | 7,5 | 128,9 | 6,1 | 8,9 | < 0.001 | TRUE |
| Multilateral collaboration | Medicine | UK | 39,8 | 46,9 | 7,1 | 17,9 | 5,7 | 8,5 | < 0.001 | TRUE |
| Multilateral collaboration | Medicine | Australia | 13,1 | 19,7 | 6,6 | 50,6 | 5,2 | 8,0 | < 0.001 | TRUE |
| Bilateral collaboration | Computer science | China | 6,4 | 13,0 | 6,6 | 102,3 | 5,2 | 8,0 | < 0.001 | TRUE |
| Multilateral collaboration | Medicine | EU-13 | 19,8 | 26,2 | 6,5 | 32,6 | 5,1 | 7,8 | < 0.001 | TRUE |

Table 3 Top 10 decreases in collaboration rate, 2010-2024

| Type | Discipline | Country | 2010 (%) | 2024 (%) | Δ (pp) | Δ (%) | 95% CI lower | 95% CI upper | p_value | Significant |
|---|---|---|---|---|---|---|---|---|---|---|
| Multilateral collaboration | Physics & astronomy | Russia | 19,6 | 10,1 | -9,5 | -48,5 | -10,9 | -8,1 | < 0.001 | TRUE |
| Bilateral collaboration | Social sciences & humanities | USA | 16,1 | 7,0 | -9,2 | -56,8 | -10,6 | -7,8 | < 0.001 | TRUE |
| Bilateral collaboration | Medicine | USA | 20,4 | 11,7 | -8,7 | -42,7 | -10,1 | -7,3 | < 0.001 | TRUE |
| Bilateral collaboration | Life & environmental sciences | USA | 15,8 | 7,5 | -8,3 | -52,4 | -9,7 | -6,9 | < 0.001 | TRUE |
| Bilateral collaboration | Engineering | USA | 11,8 | 4,0 | -7,8 | -65,9 | -9,2 | -6,4 | < 0.001 | TRUE |
| Multilateral collaboration | Engineering | EU-14 | 66,0 | 58,8 | -7,2 | -10,9 | -8,6 | -5,8 | < 0.001 | TRUE |
| Multilateral collaboration | Engineering | USA | 29,1 | 22,2 | -6,8 | -23,6 | -8,2 | -5,4 | < 0.001 | TRUE |
| Multilateral collaboration | Social sciences & humanities | USA | 37,4 | 30,9 | -6,5 | -17,4 | -7,9 | -5,1 | < 0.001 | TRUE |
| Bilateral collaboration | Computer science | USA | 14,3 | 7,9 | -6,4 | -44,7 | -7,8 | -5,0 | < 0.001 | TRUE |
| Multilateral collaboration | Computer science | EU-14 | 63,8 | 57,5 | -6,3 | -9,8 | -7,7 | -4,9 | < 0.001 | TRUE |

Table 4. Top 10 increases in RIC, 2010-2024

| Type | Discipline | Country | 2010 (%) | 2024 (%) | Δ (RIC) | Δ (%) | 95% CI lower | 95% CI upper | p_value | Significant |
|---|---|---|---|---|---|---|---|---|---|---|
| Multilateral | Engineering | Switzerland | 1,6 | 2,9 | 1,2 | 76,4 | 1 | 1,5 | < 0.001 | TRUE |

| collaboration | | | | | | | | | | |
|---|---|---|---|---|---|---|---|---|---|---|
| Multilateral collaboration | Computer science | Switzerland | 1,5 | 2,6 | 1 | 67 | 0,8 | 1,3 | < 0.001 | TRUE |
| Multilateral collaboration | Medicine | Japan | 0,9 | 1,7 | 0,8 | 84,3 | 0,6 | 1 | < 0.001 | TRUE |
| Multilateral collaboration | Life & environmental sciences | Switzerland | 1,5 | 2,3 | 0,8 | 49,3 | 0,5 | 1 | < 0.001 | TRUE |
| Multilateral collaboration | Social sciences & humanities | Switzerland | 1,4 | 2,1 | 0,7 | 53,2 | 0,5 | 1 | < 0.001 | TRUE |
| Multilateral collaboration | Medicine | Russia | 1,3 | 2 | 0,7 | 54,3 | 0,5 | 0,9 | < 0.001 | TRUE |
| Multilateral collaboration | Medicine | South Korea | 0,8 | 1,5 | 0,7 | 88,7 | 0,5 | 0,9 | < 0.001 | TRUE |
| Multilateral collaboration | Social sciences & humanities | Japan | 0,9 | 1,5 | 0,6 | 70,1 | 0,4 | 0,8 | < 0.001 | TRUE |
| Multilateral collaboration | Computer science | Norway | 1,5 | 2,1 | 0,6 | 41,9 | 0,4 | 0,8 | < 0.001 | TRUE |
| Multilateral collaboration | Engineering | Norway | 1,5 | 2,1 | 0,6 | 40,4 | 0,4 | 0,8 | < 0.001 | TRUE |

Table 5. Top 10 decreases in RIC, 2010-2024

| Multilateral collaboration | Physics & astronomy | India | 1,7 | 1,0 | -0,7 | -39,4 | -0,87928 | -0,45294 | < 0.001 | TRUE |
|---|---|---|---|---|---|---|---|---|---|---|
| Bilateral collaboration | Computer science | Russia | 0,9 | 0,5 | -0,4 | -44,1 | -0,59317 | -0,16683 | < 0.001 | TRUE |
| Multilateral collaboration | Physics & astronomy | China | 1,0 | 0,8 | -0,2 | -22,5 | -0,44817 | -0,02183 | 0.03 | TRUE |
| Bilateral collaboration | Computer science | Australia | 0,4 | 0,2 | -0,2 | -58,9 | -0,43039 | -0,00405 | 0.05 | TRUE |
| Multilateral collaboration | Engineering | India | 0,5 | 0,3 | -0,2 | -42,4 | -0,42817 | -0,00183 | 0.05 | TRUE |
| Bilateral collaboration | Engineering | Russia | 0,9 | 0,7 | -0,2 | -24,3 | -0,42539 | 0,000946 | 0.05 | FALSE |
| Multilateral collaboration | Engineering | Russia | 1,0 | 0,8 | -0,2 | -19,8 | -0,40372 | 0,022613 | 0.08 | FALSE |
| Multilateral collaboration | Computer science | Russia | 1,0 | 0,8 | -0,2 | -19,1 | -0,39706 | 0,02928 | 0.09 | FALSE |
| Bilateral collaboration | Social sciences & humanities | Russia | 0,6 | 0,4 | -0,2 | -28,6 | -0,38682 | 0,043753 | 0.12 | FALSE |
| Bilateral collaboration | Engineering | Australia | 0,3 | 0,2 | -0,1 | -38,7 | -0,33206 | 0,09428 | 0.27 | FALSE |
| Multilateral collaboration | Physics & astronomy | India | 1,7 | 1,0 | -0,7 | -39,4 | -0,87928 | -0,45294 | < 0.001 | TRUE |